\PassOptionsToPackage{hyphens}{url}
\documentclass[letterpaper,twocolumn,10pt]{article}
\usepackage{usenix-2020-09}
\usepackage{titlesec}
\titlespacing*{\paragraph}{0pt}{3pt plus 3pt}{2ex}
\titlespacing*{\subsubsection}{0pt}{2ex plus 1ex minus 0.5ex}{1ex plus 0.3ex minus 0.3ex}
\usepackage{enumitem}
\setlist{topsep=3pt plus 3pt, itemsep=3pt plus 3pt, leftmargin=2em, partopsep=3pt plus 3pt, parsep=0pt}

\usepackage{wrapfig}

\usepackage{pgfplots}
\usepackage{xcolor}
\usepackage{tikz,dblfloatfix}

\usepackage{msc}
\usepackage{authblk}
\usepackage{censor}
\StopCensoring

\usepackage{tcolorbox}

\newcommand{\roundframe}[1]{{\setlength\fboxrule{0pt}\fbox{\tcbox[colframe=black,colback=white,shrink tight,boxrule=0.5pt,extrude by=2.5pt]{\small #1}}}}

\newcommand{\RQ}[1]{\raisebox{-1pt}{\roundframe{{RQ}#1}}}

\AtBeginDocument{%
  \providecommand\BibTeX{{%
    \normalfont B\kern-0.5em{\scshape i\kern-0.25em b}\kern-0.8em\TeX}}}

\usepackage{hyperref}
\hypersetup{hidelinks}
\titlespacing*{\paragraph}{0pt}{0.8ex plus 1.2ex}{2ex}

\makeatletter
\def\putlookup#1#2{\expandafter\def\csname @lookup@#1(#2)\endcsname}
\def\lookup#1#2{\csname @lookup@#1(#2)\endcsname}
\def\iflookup#1#2{\ifcsname @lookup@#1(#2)\endcsname\expandafter\@firstoftwo\else\expandafter\@secondoftwo\fi}
\makeatother
\usepackage{color}
\usepackage{colortbl}
\usepackage{multirow,booktabs}
\usepackage{xspace,xstring}
\xspaceaddexceptions{.} 
\xspaceaddexceptions{,}
\xspaceaddexceptions{+}

\usepackage{pgfplots}
\pgfplotsset{compat=newest}
\usepgfplotslibrary{dateplot}
\pgfplotsset{compat=1.8}

\definecolor{deprecated}{RGB}{255,204,203}

\newcommand{\formatdhgroup}[3]{%
    \IfEq{#3}{MODP}{%
        DH#1\textsuperscript{#2}\xspace
    }{%
        DH#1\rlap{\textsuperscript{#2}}\textsubscript{#3}\xspace%
    }%
}

\def\dh#1#2{\IfInteger{#2}{%
    \expandafter\lookup{dhgroups}{#1#2}%
    }{%
    \expandafter\lookup{dhgroups}{0#1}\noexpand#2%
    }%
}%

\newrobustcmd{\DHcmd}[1]{%
        \expandafter\lookup{dhgroups}{#1}%
}%

\putlookup{dhgroups}{01}{\formatdhgroup{1}{768}{MODP}}
\putlookup{dhgroups}{02}{\formatdhgroup{2}{1024}{MODP}}
\putlookup{dhgroups}{05}{\formatdhgroup{5}{1536}{MODP}}
\putlookup{dhgroups}{14}{\formatdhgroup{14}{2048}{MODP}}
\putlookup{dhgroups}{15}{\formatdhgroup{15}{3072}{MODP}}
\putlookup{dhgroups}{16}{\formatdhgroup{16}{4096}{MODP}}
\putlookup{dhgroups}{17}{\formatdhgroup{17}{6144}{MODP}}
\putlookup{dhgroups}{18}{\formatdhgroup{18}{8192}{MODP}}
\putlookup{dhgroups}{25}{\formatdhgroup{25}{192}{ECP}}
\putlookup{dhgroups}{26}{\formatdhgroup{26}{224}{ECP}}
\putlookup{dhgroups}{19}{\formatdhgroup{19}{256}{ECP}}
\putlookup{dhgroups}{20}{\formatdhgroup{20}{384}{ECP}}
\putlookup{dhgroups}{21}{\formatdhgroup{21}{512}{ECP}}
\putlookup{dhgroups}{31}{\formatdhgroup{31}{X25519}{}}

\newcommand{\dhbits}[1]{\lookup{dhgroupsbits}{\expandafter#1}}

\putlookup{dhgroupsbits}{768}{\formatdhgroup{1}{768}{MODP}}
\putlookup{dhgroupsbits}{1024}{\formatdhgroup{2}{1024}{MODP}}
\putlookup{dhgroupsbits}{1536}{\formatdhgroup{5}{1536}{MODP}}
\putlookup{dhgroupsbits}{2048}{\formatdhgroup{14}{2048}{MODP}}
\putlookup{dhgroupsbits}{3072}{\formatdhgroup{15}{3072}{MODP}}
\putlookup{dhgroupsbits}{4096}{\formatdhgroup{16}{4096}{MODP}}
\putlookup{dhgroupsbits}{6144}{\formatdhgroup{17}{6144}{MODP}}
\putlookup{dhgroupsbits}{8192}{\formatdhgroup{18}{8192}{MODP}}
\putlookup{dhgroupsbits}{192}{\formatdhgroup{25}{192}{ECP}}
\putlookup{dhgroupsbits}{224}{\formatdhgroup{26}{224}{ECP}}
\putlookup{dhgroupsbits}{256}{\formatdhgroup{19}{256}{ECP}}
\putlookup{dhgroupsbits}{384}{\formatdhgroup{20}{384}{ECP}}
\putlookup{dhgroupsbits}{512}{\formatdhgroup{21}{512}{ECP}}
\putlookup{dhgroupsbits}{X25519}{\formatdhgroup{31}{X25519}{}}

\begin{document}
\pagenumbering{arabic}

\title{\Large \bf Old Habits Die Hard: Deprecated Key Exchange in Commercial VoWiFi Deployments}

\title{\Large \bf Never Gonna Give You Up: Exploring Deprecated Key Exchange in Commercial VoWiFi Deployments}

\title{\Large \bf Exploring Key Exchange Weaknesses in Commercial VoWiFi Deployments}

\title{\Large \bf Diffie-Hellman Picture Show:\\Key Exchange Stories from Commercial VoWiFi Deployments}

\makeatletter
\renewcommand\AB@affilsepx{, \protect\Affilfont}
\makeatother

\author[1,2]{Gabriel K. Gegenhuber}
\author[1,2]{Florian Holzbauer}
\author[3]{Philipp É. Frenzel}
\author[1,4]{\authorcr Edgar Weippl}
\author[5]{Adrian Dabrowski}

\affil[1]{University of Vienna, Faculty of Computer Science}
\affil[2]{UniVie Doctoral School Computer Science}
\affil[3]{SBA~Research}
\affil[4]{Christian Doppler Laboratory for Security and Quality Improvement in the Production System Lifecycle (CDL-SQI)}
\affil[5]{CISPA~Helmholtz~Center~for~Information~Security}

\def\volte{VoLTE\xspace}
\def\vowifi{VoWiFi\xspace}
\def\wifi{Wi-Fi\xspace}
\def\epdg{ePDG\xspace}

\newcommand{\todo}[1][TODO]{\colorbox{black}{\textcolor{white}{#1}}}

\newcommand{\update}[1][UPDATE]{\colorbox{black!50!red}{\textcolor{white}{#1}}}

\maketitle

\begin{abstract}

Voice over Wi-Fi (VoWiFi) uses a series of IPsec tunnels to deliver IP-based telephony from the subscriber's phone (User Equipment, UE) into the Mobile Network Operator's (MNO) core network via an Internet-facing endpoint, the Evolved Packet Data Gateway (ePDG). IPsec tunnels are set up in phases. The first phase negotiates the cryptographic algorithm and parameters and performs a key exchange via the Internet Key Exchange protocol, while the second phase (protected by the above-established encryption) performs the authentication. An insecure key exchange would jeopardize the later stages and the data's security and confidentiality.

In this paper, we analyze the \textit{phase 1} settings and implementations as they are found in phones as well as in commercially deployed networks worldwide.
On the UE side, we identified a recent 5G baseband chipset from a major manufacturer that allows for fallback to weak, unannounced modes and verified it experimentally.
On the MNO side --among others-- we identified 13 operators (totaling an estimated 140 million subscribers) on three continents that %
all use the same globally static set of ten private keys, serving them at random. 
Those \textit{not-so-}private keys allow the decryption of the shared keys of every \vowifi user of all those operators.
All these operators deployed their core network from one common manufacturer.

\end{abstract}

\section{Introduction}
\label{sec:introduction}

The term \textit{non-3GPP Access Networks} refers to the method of accessing cellular network core services without the use of a GSM/GPRS/UMTS/LTE/NR radio access network. This technique has been around since the times of GSM and has been updated multiple times since then. Some operators in the U.S. and Japan have used it to offload traffic via unlicensed Wi-Fi bands.

There are two types of non-3GPP access networks: \textit{trusted networks} (e.g., provider-operated Wi-Fi access points) and \textit{untrusted networks} (third-party Wi-Fi and Internet connections). 
In recent years, the latter variant started enjoying massive adoption as \textit{Voice over Wi-Fi} (VoWiFi), also called \textit{Wi-Fi Calling} or \textit{Voice over WLAN} (VoWLAN). For the end user, it often provides better coverage, and for the operator, it provides a way to externalize the last mile's costs while keeping the full revenue. 

On iPhone and Android, by default, VoWiFi is the preferred call termination channel when available.

At its core, \textit{untrusted non-3GPP access} works by setting up at least one IPsec tunnel to the operator's Evolved Packet Data Gateway (ePDG). It uses the Internet Key Exchange (IKE) protocol~\cite{rfc7296} and relies heavily on predefined Diffie-Hellman (DH) groups, some of which are known to be weak. 
For example, since 2015~\cite{adrian15weakdh}, \dh01{}\textsuperscript{\,bits} is assumed to be breakable by motivated academic actors, while \dh02 is within reach of nation-states. 

\begin{figure}[t]
    \centering
    \includegraphics[page=1,width=1\linewidth]{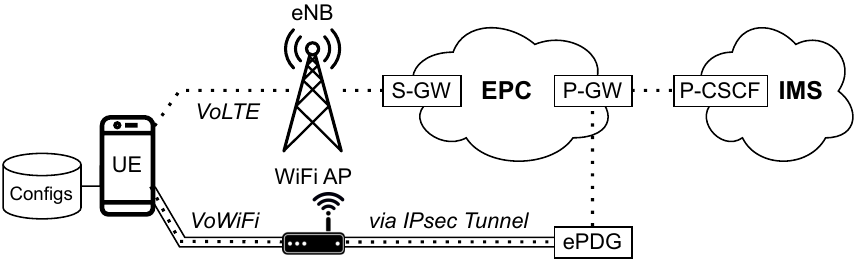}\vspace{-1mm}
    \caption{VoLTE compared to VoWiFi over an untrusted Internet connection -- as relevant for this paper}%
    \label{fig:VoWifi-overview}%
    \vspace{-2mm}
\end{figure}

\begin{figure*}[b]
    \centering%
    \includegraphics[width=0.70\linewidth]{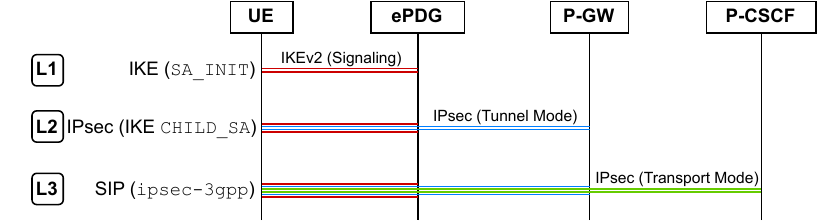}%
    \vspace{-2mm}%
    \caption{VoWiFi uses multiple tunnels to ensure security: \roundframe{L1} provides a trusted channel and manages the subsequent connections, \roundframe{L2} acts as a gateway to the internal infrastructure and \roundframe{L3} is used for the actual voice and messaging functionalities.}
    \label{fig:overview-layers}
\end{figure*}

Within the IPsec tunnel, all core network access %
is handled like regular Voice over LTE (VoLTE). Because of its recently massively increased popularity and its security perimeter function to many Evolved Packet Core (EPC) services, we investigated its specified as well as practical security from different vantage points. We facilitate static configuration analysis, active measurements of real-world implementations in network operators, as well as active measurements of handset implementations to answer the following research questions. 
\begin{enumerate}%
    \item[\RQ1] What VoWiFi key exchange methods and security parameters are preset in phones for their Mobile Network Operator (MNO)? 
    \item[\RQ2] What key exchange methods do operators actually support on their ePDG, and will they always prefer the strongest one?
    \item[\RQ3] How strong are VoWiFi connections in the real world, and how realistic is it to downgrade to weaker, breakable key exchange methods?
\end{enumerate}

We found that most operators are non-compliant with 3GPP's specifications, by still announcing and supporting deprecated DH groups weaker than 2048 bits.
Furthermore, only 42\% will take the extra step to request an upgrade if the client chooses a weaker group, but both parties actually support stronger groups.
We also found one handset manufacturer that will silently support the much weaker \dh1 group, albeit not proposing it in the handshake. \dh1 was never part of a 3GPP specification, making those handsets susceptible to man-in-the-middle attacks. We simulate such an attack by intercepting and rewriting actual \vowifi traffic.

More abstractly, our findings illustrate that functional over-provisioning and missing predefined procedures for deprecating cryptographic algorithms create a massive technical debt. %
Last but not least, we uncovered at least 13 operators\footnote{12 during our initial scan and one more during responsible disclosure.} that used the same private keys on three continents.

The paper is structured as follows. 
In Section \ref{sec:background}, we give the necessary background of how IPsec with IKE is used and embedded within the 3GPP structure. 
The threat model and the methodology are outlined in Sections \ref{sec:threatmodel} and \ref{sec:methodology}, the latter of which also includes ethical considerations.
Sections \ref{sec:staticUEconfig} to \ref{sec:downgrading} describe our implementation and report the findings, followed by an outline on how to put those findings to work for a full stack \vowifi attack. 
A related work section, a discussion, and recommendations round up the picture in Sections 
\ref{sec:relatedwork} through \ref{sec:recommendations}. %
The paper ends with a conclusion in Section \ref{sec:conclusion} and an Appendix for supplementary material.

\section{Background}
\label{sec:background}

VoWiFi is a technology that transfers voice traffic over non-3GPP access networks, typically unsecured Wi-Fi networks. It effectively routes VoLTE traffic to the EPC (and ultimately to the IMS) by encapsulating it in an IPsec tunnel over the public Internet, as shown in Figure \ref{fig:VoWifi-overview}. This basic technique has been around since the GSM era for network traffic off-loading and is now experiencing a resurgence due to the popularity of VoWiFi.

\subsection{The IKE/IPsec/SIP Stack}
The complete stack consists of a nested stack of tunnels (Figure~\ref{fig:overview-layers}). 
The outer (or \textit{Phase 1}) IKEv2 layer (\roundframe{L1} in Figure~\ref{fig:overview-layers}) is responsible for securing the inner layers (e.g., negotiating security parameters and creating key material for the nested tunnels via IKEv2~\cite{rfc5996}).
Within this layer, the client (UE) authenticates the user via the (U)SIM card and creates a CHILD\_SA (\textit{Phase 2}), that allocates an IPsec tunnel into the EPC via the Packet Gateway (P-GW).
Via this tunnel (\roundframe{L2} in Figure~\ref{fig:overview-layers}), the UE is assigned a dedicated IP address and can reach internal endpoints within the EPC.
This level of access (and also the assigned IP address) is functionally identical to connecting to the \texttt{IMS} APN over the regular radio access network (VoLTE).
Lastly, to be able to terminate voice calls, the UE uses the created CHILD\_SA (IPSec tunnel) to talk to the P-CSCF (Proxy Call Session Control Function) and establish a SIP (Session Initiation Protocol) and an RTP (Real-time Transport Protocol) connection over \textit{ipsec-3gpp}~\cite{rfc3329}, secured via IPsec in transport mode.
The encryption on this final layer (\roundframe{L3} in Figure~\ref{fig:overview-layers}) is, however, often optional and not enforced by many clients or servers. %

\subsection{Creating the ePDG Connection \roundframe{L1}}
\label{sec:bgIPsecEPC}
In the first step, the phone connects to the Internet-facing side of the ePDG server of the appropriate MNO using its Fully Qualified Domain Name (FQDN), standardized in ETSI/3GPP TS 23.003~\cite{EtsiNumberingAdressingIdentification}:\\
\indent\texttt{epdg.epc.mnc}\textit{\textlangle{}id\textrangle}\texttt{.mcc}\textit{\textlangle{}id\textrangle}\texttt{.pub.\linebreak[2]3gppnetwork.org}, where the Mobile Country Code (MCC) and the Mobile Network Code (MNC) are globally unique for each operator.

\noindent
The IKE protocol (nowadays IKEv2~\cite{rfc4306, rfc5996, rfc7296} or, more precisely, its slightly modified 3GPP variant\cite{etsi-ts-133.402}) is used to negotiate a session key using the Diffie-Hellman key exchange mechanism.
Hereby, the client proposes its supported Security Associations (SAs), i.e., the available encryption, integrity, and key exchange algorithms (DH groups) and its preferred DH group. The ePDG chooses a subset of the proposed SAs and either accepts the favored DH group or indicates its preference toward a different DH group from the proposal.
After this initial \texttt{SA\_INIT} phase, all subsequent messages are encrypted and integrity-protected.

\subsection{IPSec Tunnel Mode (CHILD\_SA) \roundframe{L2}}
After establishing the encryption on the outer IKE layer, both endpoints (i.e., the UE and the ePDG) authenticate using EAP-AKA using credentials from the (U)SIM.
Furthermore, the AKA procedure provides both parties with secret keys for the first CHILD\_SA (i.e., the IPSec tunnel into the EPC).
Note that the secret keys (and other SA parameters) used by the parent (i.e., the outer IKE) and child SAs are regularly renewed via repeated DH key exchanges and thus only valid for a certain period. However, the authentication of both endpoints is not renewed.
Thus, cracking the outer key exchange is enough to gain stealth rewriting capabilities within the first two layers.

\begin{table}[b]%
    \caption{Relevant DH groups for this work, as named/numbered by IANA \cite{iana-ike-dhgroups}}%
    \label{tab:dhgroups}%
    \begin{small}%
    \begin{minipage}[t][][b]{0.42\linewidth}%
        \begin{tabular}{c l l }
        \toprule
            Name & Bits & Type \\
        \midrule    
            \cellcolor{deprecated}DH1\footnotemark[1]&768&MODP \\
            \cellcolor{deprecated}DH2&1024&MODP \\
            \cellcolor{deprecated}DH5\footnotemark[1]&1536&MODP \\
            DH14&2048&MODP \\
            DH15&3072&MODP \\
            DH16&4096&MODP \\
            DH17&6144&MODP \\
            DH18&8192&MODP \\
        \bottomrule
        \end{tabular}
    \end{minipage}\hspace{5mm}
    \begin{minipage}[t][][b]{0.44\linewidth}
        \hspace{4mm}\begin{tabular}{ c l l }
        \toprule
            Name & Bits & Type \\
        \midrule  
            DH25 & 192 & ECP \\
            DH26 & 224 & ECP \\
            DH19 & 256 & ECP \\
            DH20 & 384 & ECP \\
            DH21 & 512 & ECP \\
            DH31 & \multicolumn{2}{l}{Curve25519} \\
        \bottomrule
        \end{tabular}
        
         \footnotesize 
        \vspace{0pt}\hfill\newline
        \footnotemark[1]never specified for 3GPP usage%
        \hfill%
          \smash{\colorbox{deprecated}{deprecated \cite{etsi-ts-133.210}}}
    \end{minipage}
       \end{small}
\end{table}

\subsection{Session Initiation Protocol (SIP) Layer}
The \textit{ipsec-3gpp} protocol that secures this layer can ensure the confidentiality and integrity of the SIP and RTP traffic.
The first two packets before establishing the encrypted channel are transmitted in plaintext (a \textit{SIP REGISTER} that is usually answered by a \textit{SIP Unauthorized} packet with the AKA challenge).
In the past, some implementations were vulnerable on the SIP layer, e.g., Exynos~\cite{Google2023Exynos}.

However, in practice, not many operators enforce encryption and integrity on this layer. In Appendix \ref{sec:SIPencryptionOptional} , we verify this experimentally. %
In such cases, an attacker who cracked the outer IKEv2 key exchange and is thus able to take over the first two layers could subsequently also hijack the third layer after the SIP authenticated between UE and P-CSCF is finished, effectively seizing control of all three communication layers. %

\begin{figure}[!t]%
\centering%
\includegraphics[clip=true,trim=3mm 10mm 48mm 22mm,width=\linewidth]{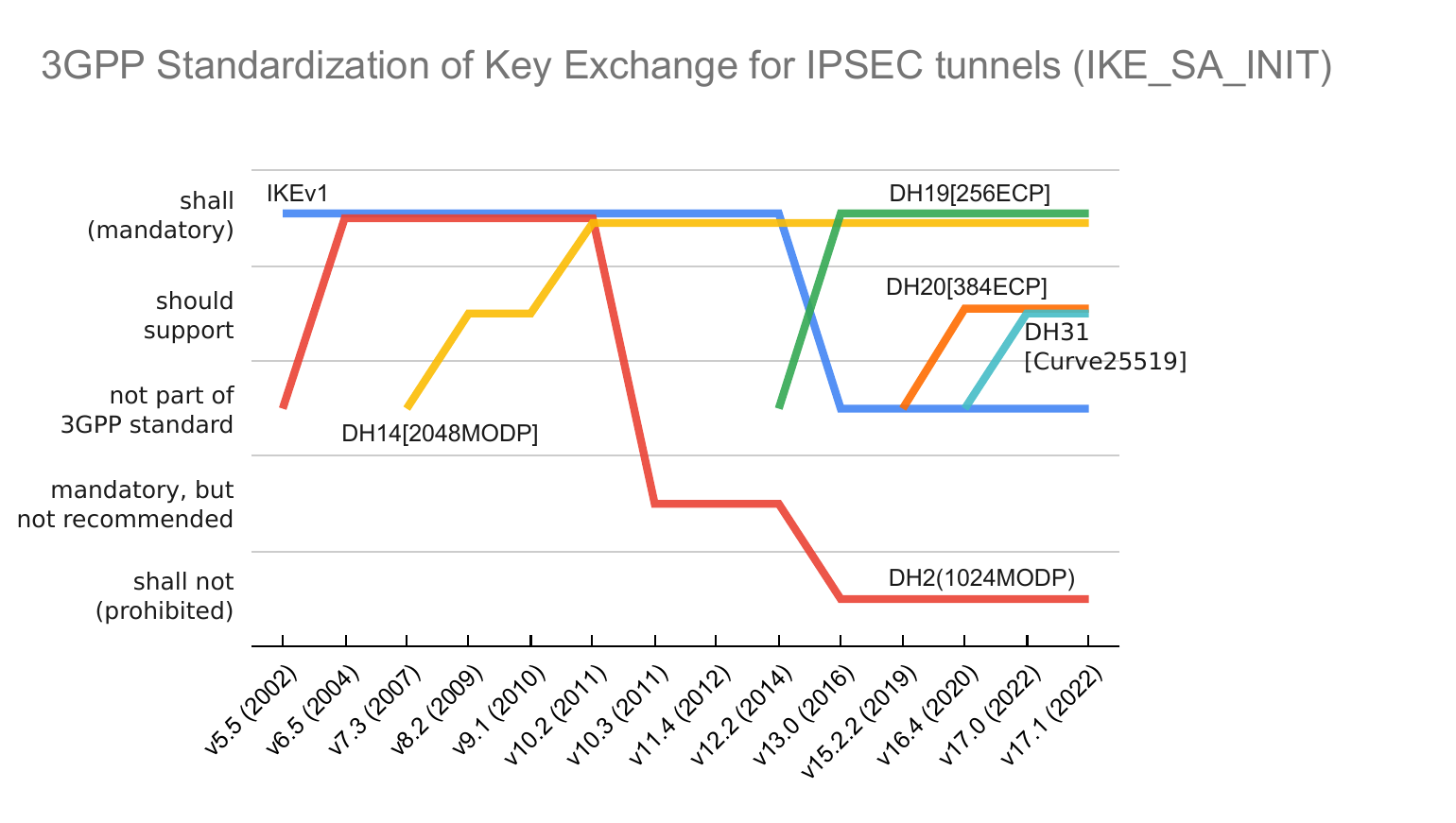}%
\caption{Development of the IPsec IKE Profile as defined in IETF TS 133.210 (IKE\_SA\_INIT) \cite{etsi-ts-133.210} Note: LTE started with v8, and IKEv1 has been phased out since v12.}%
\label{fig:dhhistory}%
\end{figure}

\subsection{IPsec/IKE Diffie-Hellman Key Exchange in the 3GPP's VoWiFi Ecosystem}
\label{sec:bgIKEin3GPP}
For most of the paper, we only look at the first-stage IKE handshake, i.e., the outermost layer \roundframe{L1}. All subsequent operations rely on the confidentiality and integrity of the negotiated encryption with the negotiated shared session key.

In contrast to much of the VoLTE/VoWiFi world (see Section \ref{sec:bgProvisioning}), IKE itself offers an automatic negotiation for key exchange mechanisms via capabilities announcements and the selection of different Diffie-Hellman (DH) groups. A DH group defines an algorithm, a key length, and a set of public parameters.
The relevant DH groups for VoWiFi (and this paper) are listed in Table \ref{tab:dhgroups}. %
In 2015, researchers estimated that cracking \dh01 %
is within the capabilities of a determined academic group, while \dh02 is within reach for nation-states \cite{adrian15weakdh}. 

Accordingly, ETSI/3GPP changed its recommendations and requirements over the years. The results of our requirement analysis for the IKE profile in TS 133 210~\cite{etsi-ts-133.210} are depicted in Figure \ref{fig:dhhistory}. While \dh01 was never part of the standard, \dh02 was recommended until 2011, then demoted to \textit{required but not recommended} and finally prohibited (\textit{shall not use}) in 2016. At the same time, new ECP\footnote{Elliptic Curve Groups modulo a Prime}-based DH groups were added as recommendations (our results show that they are rarely used).

\subsection{Modular Exponential (MODP) Diffie-Hellman Key Exchange in IKE}
\label{sec:background_dhkex}
After the client and the server agree on the key parameters, including the DH group, the server initiates a DH exchange.

Let $a$ and $A$ be the private and the public key of the server, and likewise $b$ and $B$ the private and public key of the client. Further, let $p$ be a publicly known prime and $g$ an integer smaller than $p$. $p$ and $g$ are predefined by the chosen DH group. 

The server provides the client with its public key \mbox{$A=g^a \bmod p$} along with the chosen DH group.
With that information, the client can compute its public key \mbox{$B=g^b \bmod p$} and transmit it to the server. Both parties can now compute a shared session key using \mbox{$K=B^a \bmod p$} (on the server side) and \mbox{$K=A^b \bmod p$} (on the client side).

Only $A$, $B$, and the DH group (defining $p,g$) are transmitted in clear over the wire. Ultimately, both parties know the secret symmetric session key $K$. If at least one of the private keys (or the nonce) is a fresh random integer, the generated session key $K$ will not repeat. 

\subsubsection{Optimization through Precomputation}
\label{sec:DHprecompute}
A server can precompute $A$ from a (temporarily) fixed private key $a$ for each DH group, as it is independent of the client.
This is a valid approach if the rekeying period is considerably less time than it takes a potent attacker to
crack those keys, and if the client doesn't follow a similar strategy. 

However, as pointed out by Flesh~et~al.~\cite{felsch2018dangers}, those $a$ keys should not be shared between different DH groups. Otherwise, an attacker could crack $a$ on a weak DH group and use it for stronger ones.

\subsection{VoWiFi Provisioning Ecosystem}
\label{sec:bgProvisioning}
The 3GPP VoLTE/VoWiFi ecosystem lacks a comprehensive autoconfiguration or provisioning protocol (similar to USIM files or MIB/SIB announcements used for other cellular parameters).  
This has caused (and still causes) massive compatibility problems for operating VoLTE on handsets \cite{rood2020revolte}. 
The modem and mobile OS vendors help themselves by preloading configuration databases for known MNOs in their firmware and OS images. Those configurations define a multitude of properties, from the bearer and tunnel settings down to IMS/SIP codec parameters. Some operators use an app with operator privileges to push a configuration onto the device.

The GSM Association (GSMA) approaches the problem in three ways. First, it created a database\footnote{https://imeidb.gsma.com/nsx/index} as a paid service for use by manufacturers. Second, it created a small set of standard configurations (e.g., to ease VoLTE roaming).
Third, they recently started a new Internet-based configuration service under \texttt{aes.mnc}\textit{\textlangle{}id\textrangle}\texttt{.mcc}\textit{\textlangle{}id\textrangle}\texttt{.pub.3gppnetwork.org}. At the time of writing, only 66 operators registered that domain.

\subsubsection{Apple iOS}
\label{sec:bgconfigApple}

Independent of the used modem chipset (i.e., Qualcomm or Intel), Apple organizes country-specific and operator-specific configurations into \texttt{.ipcc} files (called Country Bundles and Carrier Bundles, respectively). They can be distributed via the iOS system image and system updates as well as via itunes.com. \textit{ipcc-downloader}\footnote{\url{https://github.com/mrlnc/ipcc-downloader}} extracts them from latter source.

\subsubsection{Qualcomm: Xiaomi, Oppo}
\label{sec:bgconfigQualcomm}
Qualcomm uses proprietary encoded binary \texttt{.mbn} files to load carrier-specific modem configurations (also called \texttt{MCFGs}) into their modems.
These configuration files can be extracted from the modem image (often named \texttt{NON-HLOS.bin}) that is part of the smartphone ROM.

There have been efforts from the open source community towards providing tools to inspect the loaded configuration settings of a smartphone (e.g., \textit{EfsTools}\footnote{\url{https://github.com/JohnBel/EfsTools}}) and to sideload configurations from other smartphones with similar chipsets (e.g., to enable VoLTE support on non-carrier-branded devices).
The VoWiFi-related settings are located within the \texttt{\path{/data/iwlan_s2b_config.xml}} file of the unpacked configuration tree.

\subsubsection{Samsung}
\label{sec:bgconfigSamsung}
\label{sec:samsung}
The VoWiFi configuration on Samsung devices can be found within the \texttt{\path{/system/etc/epdg_apns_conf.xml}} file on the smartphone.
We believe these settings are also used with other modem chipsets on Samsung devices since the file also exists in ROMs for MediaTek- and Qualcomm-equipped models.
In contrast to the OEMs mentioned above (e.g., Xiaomi), the Qualcomm-based Samsung devices do not contain additional \texttt{.mbn} modem configurations in their modem image.

\subsubsection{Google Pixel}
\label{sec:bgconfigPixel}
\label{sec:google-pixel}
Google Pixel generations up to the Pixel~5 used a Qualcomm chipset, utilizing the Qualcomm configuration approach described above.
Starting with the Pixel~6, Google introduced its own Tensor-based SoCs, where VoWiFi-specific configuration parameters are consolidated into Android-generic \textit{Carrier Configuration} settings\footnote{\url{https://source.android.com/docs/core/connect/carrier}}.
However, inspecting the publicly accessible operator-specific configuration files\footnote{\url{https://android.googlesource.com/platform/packages/apps/CarrierConfig/+/main/assets}} shows that the responsible \texttt{iwlan} settings are not used in practice.
Besides shipping \textit{Carrier Configurations} via the Android-wide presetting, operators can change these settings via their own carrier app (to gain \textit{Carrier Privileges}, an app needs to be signed with a specific certificate that is saved on the SIM card).
In practice, many modern Pixel phones fall back to the default values (defined in the Android source code\footnote{\url{https://android.googlesource.com/platform/frameworks/base/+/refs/heads/android14-release/telephony/java/android/telephony/CarrierConfigManager.java\#9099}}).

\section{Threat and Attacker Model}
\label{sec:threatmodel}
\paragraph{Goals}
From an adversary's perspective, three main goals motivate an attack:
\begin{itemize}\parskip=3pt plus 3pt\itemsep=0pt
    \item[\roundframe{G1}] Eavesdropping on private communications (e.g., extracting the signaling or voice channel of realized calls or spying on sent SMS messages).
    \item[\roundframe{G2}]  Using the trusted communication channel as an attack vector towards the phone (e.g., by injecting maliciously formed SIP messages as seen in the recent Exynos vulnerabilities~\cite{Google2023Exynos}).
    \item[\roundframe{G3}]  Injecting actions towards the provider (e.g., spoofing SMS messages to impersonate the user or monetizing the exploit by calling value-added numbers) or EPC access in general.
\end{itemize}

\paragraph{Capabilities}
Traffic interception and modification can happen at any point over the Wi-Fi (e.g., via ARP/RA spoofing, the Wi-Fi access point (e.g., from a hotspot operator), or while on Internet transit.

For some of the presented attacks, we further assume a determined attacker with the capability to break \dh01 or even \dh02, according to Adrian~et~al.~\cite{adrian15weakdh}.

\paragraph{Criteria}

If both the server and client support those weak DH groups and actually use them (either by tricking them or by default config), those VoWiFi tunnels into the EPC would be vulnerable. 

Further, any divergence from the privateness (secrecy) of a \textit{private key} constitutes a broken encryption.

\section{Methodology}
\label{sec:methodology}
To explore the VoWiFi landscape and answer RQ1-3, we had to approach it from multiple vantage points: 
a) We examine the different client-stored operator configurations. 
b) We test the configuration of commercial network operators worldwide and their corner cases. 
c) We examine real UE-operator interaction by observing and modifying traffic. 
In the Discussion in Secion \ref{sec:discussion}, we then contrast those results.

\subsection{Static Client-Side Configuration Analysis}%
\label{sec:methodology_clientside}
\setlength{\intextsep}{0pt}
\setlength{\columnsep}{6pt}
\begin{wrapfigure}{r}{0.6\linewidth}\vspace{-0.4em}%
    \includegraphics[page=2,width=1\linewidth]{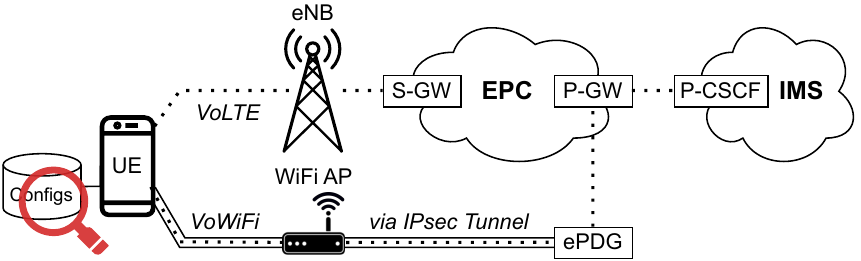}%
\end{wrapfigure}%
To approach \RQ1, we chose a static configuration analysis since otherwise, we would need a valid SIM card for each operator. 

As stated in Section \ref{sec:background}, critical information about the VoLTE and VoWiFi data bearer (or tunnels), as well as the IMS settings, need to be known to the UE in advance for each MNO.
In lieu of a 3GPP autoconfigure protocol, a database of known settings for each operator is preloaded to the device.
Due to different VoLTE and VoWiFi implementations -- depending on the OS, OEM, and modem chipset manufacturer -- there is no standardized way to access these settings. 
Thus, we extract those settings from Apple, Qualcomm-based, Tensor-based (Pixel), and Exyonos-based (Samsung) phones separately, roughly following the market share~\cite{SmartphoneMarketShare2023}.

We identified the following interesting parameters: 
 1)~the key exchange methods (e.g., Diffie--Hellman groups),
    2)~rekeying timers,
    and 3)~encryption, integrity algorithms \& pseudo-random function (PRF).

\subsection{Active MNO-Side ePDG Scanning}
\label{sec:methodology_mnoscan}
\setlength{\intextsep}{0pt}
\setlength{\columnsep}{6pt}
\begin{wrapfigure}{r}{0.6\linewidth}\vspace{-0.5em}%
    \includegraphics[page=4,width=1\linewidth]{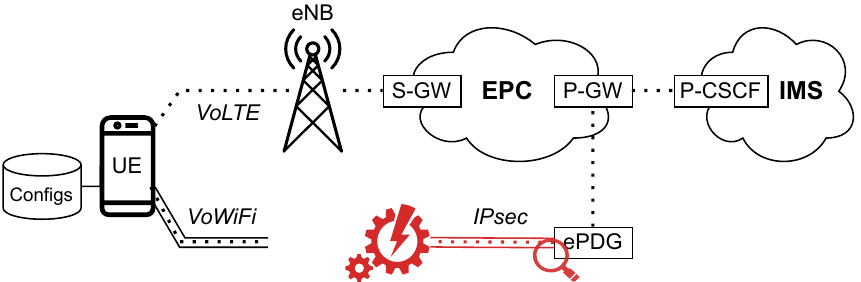}%
\end{wrapfigure}%

To answer \RQ2, we have to analyze operators' IKE handshakes and probe different key exchange methods.
First, we query all possible DNS names (described in Section \ref{sec:background}), resolving DNS requests in an iterative manner (i.e., getting all the IP addresses from authoritative servers).
For each operator, we try to negotiate an IKE phase 1 key with every key exchange method from Table~\ref{tab:dhgroups} separately -- i.e., we pretend to support only one method at a time. 
For all operators, we record how their servers react and whether they propose a different DH group (if so, which one) or accept the client's choice for a stronger one.
Additionally, we test if the server tolerates the client's choice for a weaker DH group, even if both parties announce support for a stronger one. This would ease downgrade attacks to an attackable bit length.

\subsection{Testing Implementations and Composition}
\label{sec:methodology_composition}
\setlength{\intextsep}{0pt}
\setlength{\columnsep}{6pt}
\begin{wrapfigure}{r}{0.6\linewidth}\vspace{-0.5em}%
    \includegraphics[page=3,width=1\linewidth]{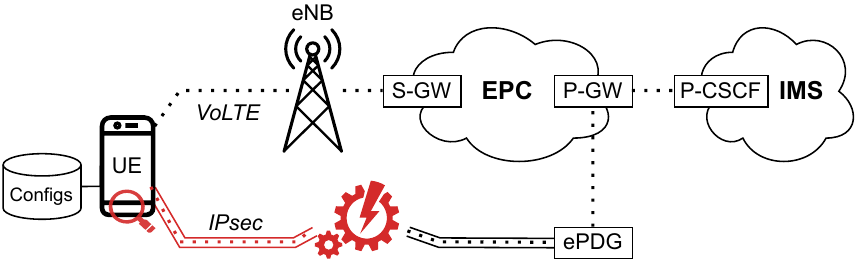}%
\end{wrapfigure}%
To answer \RQ3, we need to combine multiple results from the client and the server side to form a view of the complete system and its security properties.

\subsubsection{Downgrading Possibilities}
We examine if phones from different manufacturers accept weaker DH groups even if higher ones are available, with an active test against the phone. We also want to know if phones support any undocumented DH groups.
To this end, we redirect real traffic between our phone and the ePDG to a script that allows us to intercept and rewrite data.  

Further, we test if networks would accept a weaker DH group despite both parties indicating support for higher groups. %
To this end, we probe each operator's ePDG but announce multiple methods at once.

\subsubsection{Interception Opportunity}
When attacking the key, the attacker has to outrace the rekeying period of connections, i.e., crack the key before it loses validity. We extract those settings from the phones.
Putting both sides together, we also consider key value re-usage and undocumented DH groups.

\subsection{Limitations}
\label{sec:Methodology_Limitations}

\subsubsection{Limited MCC-MNC Mapping} 
\label{sec:MethodologyLimitationsMapping}
One operator can have multiple MCC-MNC %
designations, for example, because of past mergers. Likewise, one MCC-MNC tuple can (but does not have to) be shared between multiple virtual network operators. This depends on whether the MVNO operates its own Home Subscriber Server (HSS). MVNOs can also contract services, i.e., share the same physical servers with another operator.

We refrained from error-prone manual disambiguation. Thus, unless otherwise stated (for a very particular vulnerability), results operate on a one-$\langle$MCC-MNC$\rangle$-tuple-per-operator approximation.

\subsubsection{Consistent Configuration}
For resilience and load balancing, operators could either explicitly or invisibly operate multiple servers/gateways, some of which could have diverging configurations. Explicit load balancing is externally visible, e.g., via DNS round robin, while a dedicated load balancer would conceal load balancing from the outside world (it has only one IP address). Unless otherwise stated, we assume consistent configurations within an operator for our measurements and results.

\subsection{Ethical Considerations}
\label{sec:Methodology_Ethical}
\label{sec:Ethical}

Since some parts of this paper include measurements on real-world production provider infrastructure, we assessed its necessity and the least invasive method to perform the investigation.  

\paragraph{Invasiveness.} 
We always measured properties with our devices or in sandboxes unless the research subject required real-world data. 

\paragraph{Traffic and Server Load.}
Connections to production systems were of low volume and with the minimum number of connections necessary for the task. Since those systems (e.g., ePDG) are made to handle traffic for (millions) of customers, we are confident that our attempts did no harm.

\paragraph{Traffic Abnormality.} 
Our handshake attempts for the ePDG always confirmed the appropriate RFC format and never contained illegal data or malformed structures.

\paragraph{Confidentiality and Integrity.}
Our handshake attempts never contained real credentials and, therefore, should never have access to any privileged functions of confidential data.

\section{Static Client-Side Configuration}%
\label{sec:staticUEconfig}
To cover a considerable share of real-world client configurations, we analyzed different implementations and extracted the corresponding settings for the available operators out of smartphones and smartphone firmware images that reflect the current market situation~\cite{SmartphoneMarketShare2023}.
While not part of our threat model, we additionally extract encryption, integrity and pseudo-random function (PRF) algorithms alongside DH groups and rekey timers for the IKEv2 security association parameters and evaluate their prevalence. %

\subsection{Implementation}

After downloading and extracting the available \textbf{Apple iOS} carrier bundles (Section \ref{sec:bgconfigApple}),
we filter for iPhones (discarding other device types such as Apple watches) and group them by operator. For our statistical analysis, we select the latest VoWiFi configuration for each operator.

Configurations for \textbf{Qualcomm}-based phones, such as Xiaomi and Oppo, use the Qualcomm \texttt{.mbn} mechanism as described in Section \ref{sec:bgconfigQualcomm}. Leveraging the information from other open-source projects, we implemented a
parsing tool\footnote{\url{https://github.com/sbaresearch/mbn-mcfg-tools}}
to unpack and parse the modem configuration %
files.
We analyzed configuration files from the Xiaomi~13~Pro (2023-08-22) and the Oppo~X6~Pro (2023-12-06).

In the category of \textbf{Samsung}'s VoWiFi configurations (see Section \ref{sec:bgconfigSamsung}), we analyzed the most recent (2023-12-29) configuration file from the Exynos-based Galaxy S24+.

As \textbf{Google Pixel} phones have multiple ways to receive carrier configurations, we used them primarily to extract Android 14 default values.

\paragraph{IKEv2 Default Values}
\label{sec:default-values}
We focused our client-side analysis on operator-specific settings, overriding the default state.
However, operators may also refrain from providing specific values, leading to a fallback to a predefined default.
Additionally, these default settings are also used for operators that are not part of the preloaded configuration files at all (e.g., smaller mobile virtual network operators, MVNOs).

In our static analysis, we were able to recover the default settings for Samsung devices (cf. Section~\ref{sec:samsung})
and for newer Pixel phones (cf. Section~\ref{sec:google-pixel}).

\begin{table*}[!b]
    \vspace{-1ex}
    \centering
    \caption{Default parameters for IKEv2 if no MNO-specific configuration is present.}
    \label{tab:ike_defaults}
    \begin{footnotesize}
    \begin{tabular}{l l l l l}
        \toprule
        \multicolumn{2}{l}{Vendor} & Qualcomm (Xiaomi\textsuperscript{\textdagger}) & Samsung & Google Pixel \\ \midrule
        \multirow{5}{*}{\rotatebox{90}{Defaults}}&  DH Group & \smash{\colorbox{deprecated}{\dh02, \dh05}}, \dh14 & \smash{\colorbox{deprecated}{\dh02}} & \smash{\colorbox{deprecated}{\dh02, \dh05}}, \dh14 \\
        & Rekey Timer & 64,800s (soft), 64,900s (hard) & 86,400s & 7,200s (soft); 14,400s (hard)\\
        & Encryption & AES\_CBC\textsuperscript{128,256}, 3DES & AES\_CBC\textsuperscript{128}  & AES\_CBC\textsuperscript{128,192,256} \\
        & Integrity & SHA1\textsuperscript{96}, AES\_XCBC\textsuperscript{96}, MD5\textsuperscript{96} & SHA1\textsuperscript{96} &  XCBC\textsuperscript{96}, SHA1\textsuperscript{96}, SHA2\textsuperscript{256,384,512} \\
        & PRF & SHA2\textsuperscript{256}, SHA1, AES\textsuperscript{128} & * & SHA1, AES\_XCBC\textsuperscript{128}, SHA2\textsuperscript{256,384,512}\\
        \bottomrule
    \end{tabular}
    \begin{minipage}[c]{0.85\linewidth}\scriptsize
\vspace{1mm}
\centering
* If no PRF is set, the PRF can be derived from the integrity algorithms. \hspace{1ex}
\textsuperscript{\textdagger}~Xiaomi Poco X3 NFC \hspace{1ex} \smash{\colorbox{deprecated}{deprecated DH\cite{etsi-ts-133.210}}}
\end{minipage}%

    \end{footnotesize}
\end{table*}

\subsection{Results}

Table~\ref{tab:client_parameters} compares the presence of operator-specific VoWiFi settings in our analyzed client configuration files.
The percentage column shows the share of operators that actually provide dedicated VoWiFi settings.
Figure~\ref{fig:dhgroups} shows the prevalence of the different DH groups in the analyzed client configurations. We see that on the client side, DH groups with larger key sizes have not reached widespread support yet.
Within all analyzed device groups, only a single operator (T-Mobile Germany) on Samsung devices signals support for an elliptic curve group (i.e., \dh19).

\begin{table}[t]%
    \centering%
        \caption{IKEv2 security association parameters inside static UE configurations}%
    \label{tab:client_parameters}%
    \resizebox{\linewidth}{!}{%
    \begin{tabular}{l l r r r r}
    \toprule
    \multicolumn{2}{l}{Vendor} & Apple & Xiaomi  & Oppo  & Samsung \\
    \midrule
   \multirow{5}{*}{\rotatebox{90}{Configs}} & DH Group & 219 (29\%) & 150 (56\%) & 221 (59\%) & 156 (49\%) \\
   & Rekey Timer & 219 (29\%) & 231 (86\%) & 340 (90\%) & 95 (30\%) \\
   & Encryption & 219 (29\%) & 126 (47\%) &  211 (56\%)& 141 (44\%) \\
    & Integrity & 219 (29\%)  & 130 (48\%)  & 212 (56\%) & 141 (44\%) \\
   & PRF & 219 (29\%) & 120 (44\%) & 203 (54\%) & 0 (0\%) \\ \hline 
    \multicolumn{2}{l}{Total MNO Configs} & 745 & 270  & 377 &  319 \\
    \bottomrule
      
    \end{tabular}
    }

\end{table}

\subsubsection{Apple iOS}
\label{sec:staticResultsApple}
Of a total of 745 operator-specific \texttt{.ipcc}-configurations for iPhone devices, 219 specify VoWiFi-related settings.
The remaining 526 operators either do not support VoWiFi yet or rely on the device's default configuration.

\begin{figure}[t]
    \centering\includegraphics[width=1\linewidth,clip=true,trim=2mm 2mm 2mm 2mm]{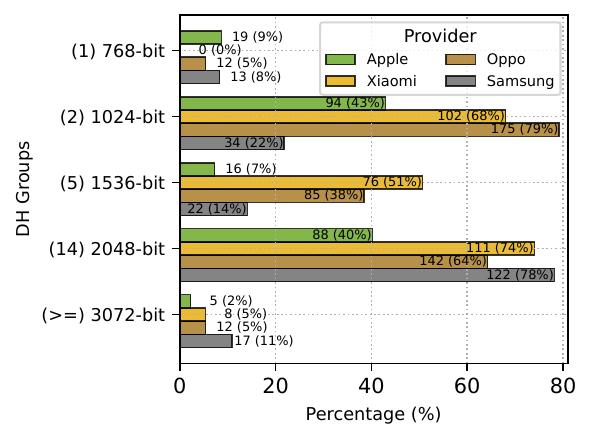}%
    \vspace{-2mm}%
    \caption{Number of MNOs per supported DH group (client side, grouped by device type).}%
    \label{fig:dhgroups}
\end{figure}

Analyzing the operator-specific VoWiFi settings for iPhones, we discover two properties:
\begin{enumerate}\itemsep=0pt\parskip=3pt plus 3pt\topsep=0pt\partopsep=0pt
    \item While other vendors (e.g., Qualcomm, Samsung) usually define a broad set of supported security parameters, Apple, with the exception of three MNO configs, only defines a single algorithm setting for each VoWiFi-related attribute. 
    Thus, on the network, it just signals support for one single DH group. 
    The same holds true for the other configuration settings (e.g., rekeying or encryption and integrity algorithms).
    \item Whenever one IKEv2-related parameter is set for an operator, the configuration also contains all the other parameters. %
    (i.e., the \texttt{.ipcc}-configurations always contain complete settings). This can be seen in Table~\ref{tab:client_parameters}, as all columns contain the same percentage values (i.e., 29\%).
\end{enumerate}
Due to these properties, Figure~\ref{fig:dhgroups} shows that iPhones exclusively support (never 3GPP-standardized) \dh1 to connect to 9\% of the analyzed operators.

\subsubsection{Android}
\label{sec:staticResultsAndroid}
Qualcomm \texttt{.mbn} files can be deployed and adjusted both by the chipset vendor and OEMs, leading to a different number of \texttt{.mbn} files for Xiaomi and Oppo.
Our analyzed Xiaomi device includes 270 \texttt{.mbn} files, compared to 377 for the selected Oppo smartphone.
While MBN files are more likely to include VoWiFi-specific settings, they do not always specify all parameters (as opposed to Apple, Section \ref{sec:staticResultsApple}).
Nearly every \texttt{.mbn} file includes a rekey timer and differentiates between soft and hard timers.
The soft timer specifies the number of seconds until the client tries to renew the corresponding SA. The hard timer states the maximum lifetime of an IKEv2 SA.
If only one timer is specified, as is the case for Samsung, it represents the hard timer, thus the total lifetime of the SA.
In contrast to the rekey timer only half of the \texttt{.mbn} files include SA parameters such as the DH groups, encryption, integrity, or PRF algorithms.

\subsubsection{Default (Fallback) Values}
As described in Section~\ref{sec:default-values}, we extracted the default values for Samsung devices and recent Google Pixel phones.
Since there were no IKEv2-specific SA parameters available within our default Qualcomm \texttt{.mbn} profiles, the used settings are taken from the modem's default (defined in even deeper layers of the modem firmware).
To gain comparable settings for Qualcomm, we thus extracted the proposed values from an active capturing of our lab's Qualcomm-based Xiaomi device (the Xiaomi Poco X3 NFC, using the Snapdragon 732G) when no specific carrier \texttt{.mbn} file was loaded.
We list the default values in Table~\ref{tab:ike_defaults}.
As the table shows, Samsung only sets one specific value for each IKEv2 parameter category (similar to the operator-specific behavior observed for iPhones).
In contrast, our Xiaomi and Google Pixel devices propose various settings to the server endpoint.

Both Xiaomi and the Pixel phone default to the first (and weakest) setting and propose a \dh2 key exchange within the first \texttt{SA\_INIT} handshake packet.
Note that the initial DH client preference is not relevant for active attackers, because it can be arbitrarily switched by sending an IKEv2 protocol extension packet to the client (as described in Section~\ref{sec:pivoting-dh-groups} towards the end of the paper).

In the first packet (\texttt{SA\_INIT}), the UE has to choose a DH group from the list. During our test, the first and weakest DH group \dh02 was chosen.

\subsubsection{Deprecated IKE Parameters}
\begin{figure}[!t]%
    \centering\includegraphics[width=1\linewidth,clip=true,trim=2mm 2mm 2mm 2mm]{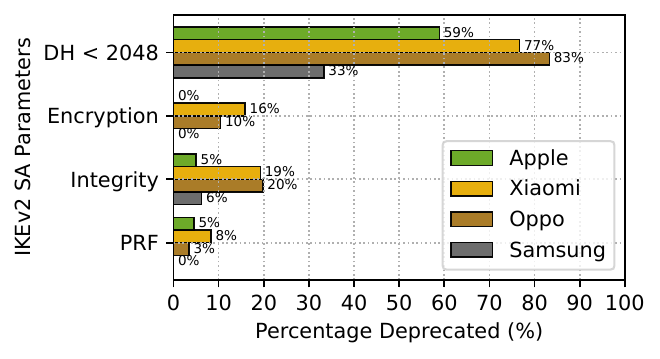}%
    \vspace{-2mm}%
    \caption{Share of deprecated IKEv2 parameters within all operator-specific VoWiFi settings, i.e., 83\% of Oppo's configured DH settings include a deprecated DH group.} %
    \label{fig:ike_deprecated}
\end{figure}

Our static analysis of IKEv2 security parameters on the client side shows an alarming share of deprecated algorithms.
Figure~\ref{fig:ike_deprecated} shows the deprecation share by each IKEv2 security algorithm group and device type.

\dh2 is the most dominant group among the deprecated groups, but also among all measured groups in total for many devices (e.g., Apple, Xiaomi, and Oppo). It is also the go-to fallback value for many configurations (cf. Table~\ref{tab:ike_defaults}).

Regarding encryption and integrity, many clients still support the deprecated DES and MD5 algorithms.
Table~\ref{tab:deprecated} in the Appendix lists decrepated IKEv2 SA algorithms.

Note that Figure~\ref{fig:ike_deprecated} only shows the results of the operator-specific settings, not considering default values.
For example, Samsung only shows a DH group deprecation of 33\%. 
However, only 49\% of the operators override the default value (cf. Table~\ref{tab:client_parameters}); thus, in practice, the deprecated \dh2 group is used as a fallback in many real life scenarios.

\subsubsection{Key Lifetimes}
\label{sec:keylifetimes}
The key lifetime on the IKEv2 layer essentially defines the available timeframe for cracking the key.
From a security perspective, shorter lifetimes (and, obviously, strong DH groups) are recommended to extend the time and resources needed to crack the key. %

\begin{figure}[t]
    \centering%
    \includegraphics[width=0.9\linewidth,clip=true,trim=0mm 2mm 0mm 2mm]{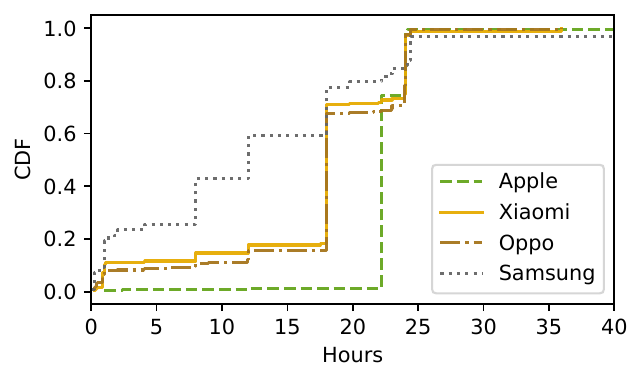}%
    \vspace{-2mm}%
    \caption{Operator-specific configuration of rekey timings. The majority of Apple devices is configured to renew the keys after 22 hours. 
    For Xiaomi and Oppo, the graph represents the configured soft timers (peaking at 18 hours).}%
    \label{fig:rekeytimer}
\end{figure}

Figure~\ref{fig:rekeytimer} shows the rekey intervals set by each vendor.
In almost all cases, re-keying takes place within a 24-hour time frame.
40\% of Samsung devices tried to rekey in the first 10 hours, while most iPhones rekey after 22 hours, which should leave enough time to be in reach for nation-state attackers.
We observed three outliers inside Samsung's MNO configurations that specify a key lifetime of a year.
The 3GPP specification~\cite{etsi-ts-133.210} does not give recommendations for rekey timers, which ultimately delegates the decision to the operators.

\subsubsection{Client Side Validation (Sanity Check)}

We used a random sample (n=12) of available smartphone devices (i.e., all testing devices from our lab and some additional models from volunteers that allowed us to record the IKEv2 handshake from their regular smartphone) to do a sanity check and verify whether the obtained results from our static analysis are feasible.
Our selection covers every device group from our static analysis with at least one model (i.e., using iPhones, Qualcomm-based devices, Samsung models, Google Pixel, and additionally, several MediaTek-based devices).
Although the extracted IKEv2 proposals from our captures are biased towards operators from our home country \censor{Austria}, we used them as a sanity check to verify the results from our static analysis.
For all devices that matched the exact models from the configuration file analysis (i.e., iPhone and Google Pixel), we were able to verify the obtained results, i.e., the proposals were identical to the settings in the configuration files.
Moreover, the residual devices from our sample also showed a similar distribution (e.g., \dh2 being the most popular DH group).

\section{Active MNO-side ePDG Scanning}
\label{sec:activemnoscanning}

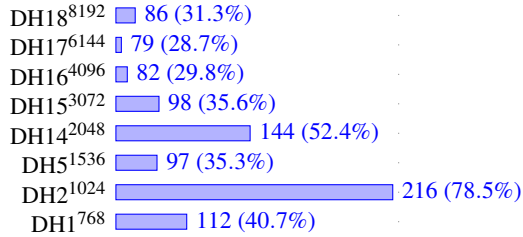
\begin{figure}[t!]
\centering
\begin{tikzpicture}
  \begin{axis}[%
    xbar,
    y axis line style = { opacity = 0 },
    axis x line       = none,
    tickwidth         = 0pt,
    ytick             = data,
    enlarge y limits  = 0.05,
    enlarge x limits  = 0.02,
    width=0.3\textwidth,
    ticklabel style = {font=\small},
    bar width=2mm,
    symbolic y coords = {768,1024,1536,2048,3072,4096,6144,8192},
    yticklabels={\dhbits{768},\dhbits{1024},\dhbits{1536},\dhbits{2048},\dhbits{3072},\dhbits{4096},\dhbits{6144},\dhbits{8192}},
    nodes near coords={\small \pgfmathprintnumber
{\pgfkeysvalueof{/data point/x}} (\pgfmathparse
{\pgfkeysvalueof{/data point/x}/275*100}\pgfmathprintnumber[fixed,precision=1]{\pgfmathresult}\%)},
  ]
  \addplot coordinates {
    ( 112,768) 
    ( 216 ,1024) 
    ( 97 ,1536)
    ( 144 ,2048) 
    ( 98 ,3072)
    ( 82 ,4096) 
    ( 79 ,6144) 
    ( 86 ,8192)
  };
  \end{axis}
\end{tikzpicture}\vspace{-1ex}%
\caption{Number of MNOs per supported DH group}%
\label{fig:dhpopularity}
\end{figure}

\subsection{Implementation}
To analyze operators' IKE handshakes and probe different key exchange methods we operated as follows.
First, we queried all possible ePDG DNS names (Section \ref{sec:bgIPsecEPC}) with \textit{massdns}\footnote{\url{https://github.com/blechschmidt/massdns}}, delegating all queries to a local \textit{unbound}\footnote{\url{https://github.com/NLnetLabs/unbound}} instance, iteratively resolving DNS requests (i.e., getting the IP addresses from the authoritative server).
Afterward, our Python-based IKEv2 implementation tried to negotiate a key with each of the methods from Table~\ref{tab:dhgroups} with every operator.
Our implementation\footnote{https://github.com/sbaresearch/vowifi-epdg-scanning} is based on predefined packet structures from \textit{scapy}\footnote{\url{https://github.com/secdev/scapy/blob/master/scapy/contrib/ikev2.py}} and was verified %
against a self-hosted \textit{strongSwan} server.
For each tested operator, we recorded the server's answer, including any optional DH group suggestions, the public key value, and additionally the whole interaction as a PCAP file. 
Additionally, we tested if the server tolerates a client's choice of a weaker DH group, even if both parties announce support for a stronger one. This would ease downgrade attacks to a feasibly attackable bit length.

\subsection{ePDG Supported Key Exchange Methods}
As of Q4 2023, operators maintained 423 \epdg domain names (minimum one A record, of which 16 additionally provided \texttt{AAAA} records).
Of these 423 operators, 275 responded to our handshake, of which 33 rejected all of our proposed key exchange methods. 
We suspect that some might have geoblocked their VoWiFi services to prevent roaming evasion or for increased security.

\subsubsection{MODP Groups}
275 ePDG servers responded to our handshake attempts. By offering only one DH group, we tested the servers' capabilities. Some servers tend to ignore requests with unsupported groups, while most reported a handshake error; none proposed a downgrade. As depicted in Figure \ref{fig:dhpopularity}, 79\% support \dh02, followed by \dh14 with  52\%, and \dh01 with 41\%.

\begin{figure}[t]
    \centering
    \includegraphics[width=0.8\linewidth]{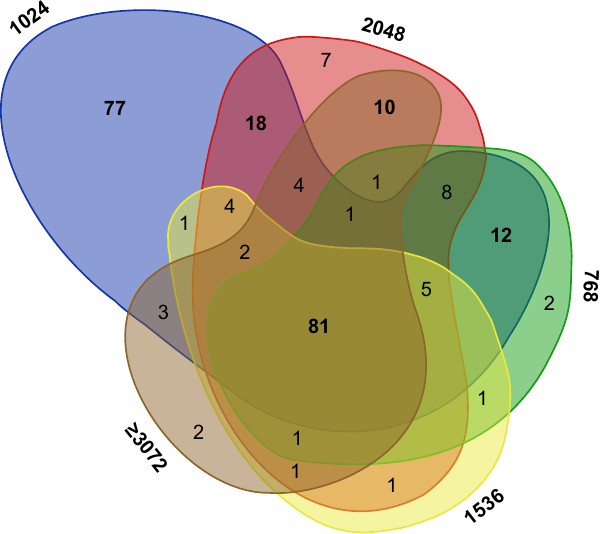}
    \caption{Number of MNOs that support a specific combination of DH key exchange groups. 3072-8192 bit groups are combined because of their low diversity.}
    \label{fig:supportedDH}
\end{figure}

Figure \ref{fig:supportedDH} shows the combinations of supported methods per operator. Only two operators solely support \dh01, and 77 only support \dh02. The former was never proposed by the 3GPP for usage \cite{etsi-ts-133.210}. 12 and 18 operators support combinations of \dh01+\dh02 and \dh02+\dhbits{2048}, respectively. 
Once an operator chooses to support \dhbits{3072}, it usually supports most of the groups up to \dhbits{8192}. 65 operators supported all groups from \dh01 up to \dhbits{8192}.

\subsubsection{ECP Groups}
Except for one private operator, there is no support for elliptic curve groups.
13 operators proposed a downgrade to \dh01 in their response, even though all of them support up to \dhbits{8192}. Two operators proposed a downgrade to \dh14. All others (including T-Mobile Germany, which signals ECP support in the client-side config on Samsung devices) either ignored the handshake, returned a negative answer, or reported an error.
Thus, in practice, ECP appears to be rarely used for real-world VoWiFi connections.

\subsection{
Tolerating Weak DH Preferences
}
\label{sec:tolerance}
We want to know if ePDGs tolerate weaker DH groups than their common set of supported methods allows for. In this test, our client connected, indicating support for all DH groups, but chose \dh02 as the preferred one. 

41\% of the operators accepted the proposed less secure method, 12\% returned an error without indicating which group to choose instead, but 42\% desired an upgrade by the client. Roughly half of them chose \dh18, most of the others \dh14, with a few single-digit outliers requesting %
\dh15, \dh16, and \dh05 (in descending order). 

Curiously, 4\% seemed to indicate a desired downgrade to \dh01. However, as all those networks actually support our proposed \dh02, this probably represents a generic \textit{make it work at all costs} error message (e.g., if some of the higher groups are not recognized, similar to what we have seen with ECP groups).

\subsection{Inter-MNO Static Key Sharing}

In our scans, initially\footnote{After the manufacturer provided a fix, a 15th ePDG/13th MNO appeared.}, 14 ePDG servers (12 operators, based on IP addresses and background story, see Appendix \ref{sec:mnocounting})  showed a very peculiar behavior: They repeatedly served the same keys. 
A repeated scan with apx. 500 \dh02 handshakes on those MNOs revealed a globally shared set of exactly ten static public keys randomly used on each connection attempt by every one of those operators.

However, this, in return, means that those 12(+1) operators all use the same ten private keys. %
Violating the secrecy requirement of the \textit{private key} allows any of those operators (or anyone else who seizes the keys from them legally or through other means), as well as the originator of those keys, to decrypt any other operators' shared session secrets instantly.%

Using the notation from Section \ref{sec:background_dhkex}, if an attacker can read the plaintext DH group and $B$ from the wire and knows one of the private keys (in our case $a$), the secret session key can be reconstructed using \mbox{$K=B^a \bmod p$}.

In a smaller sample, we also confirmed that similar sets exist for other DH groups on the same operators. %
As the key $a$ is independent of the DH group, an attacker who does not know the private key $A$ can crack the weakest group \dh01 and then use it to reconstruct $K$ generated for the stronger groups.

Using passive banner analysis with Shodan\footnote{https://www.shodan.io/}, we confirmed that at least three of those ePDGs are from ZTE (the others were firewalled). For all MNOs (except for one), press releases show contracts with, winning bids by, or strategic cooperation with ZTE to build an LTE or 5G network. 
Eight of those networks are located in Asia, three in central Europe, and two in South America.

Without knowledge about how those operators arrived at using the same static set of then non-randomized keys, we initiated a responsible disclosure with the GSMA. The process, the manufacturer's response, and a list of key hashes are to be found in Appendix \ref{sec:responsibledisclosureGSMA}. 
We later found that the same ten keys are also used for the phase 2 \roundframe{L2} key exchange.

\subsection{Intra-MNO Key Reusage}
\label{sec:intra-mno-key-reusage}
We also encountered MNOs that reused keys between handshakes. 
If handled carefully, this can be a valid optimization on the server side as described in Section \ref{sec:DHprecompute}.

We have also encountered rare instances of nonce reuse, which violates the IKEv2 specification and also defies the common definition of \textit{number used once}.

\section{Downgrading Vulnerabilities}
\label{sec:downgrading}
Based on the results from the above sections, we devise experiments to assess and test the resilience against downgrade attacks. As per our threat model from Section \ref{sec:threatmodel}, a downgrade to a sufficiently weak key exchange method is considered a successful attack. 

\subsection{Implementation}
As described in the threat model in Section~\ref{sec:threatmodel}, a user's traffic can be intercepted locally (e.g., by a malicious WiFi operator), anywhere on the path, or on a large scale (e.g., by a nation-state monitoring an IXPs traffic).
To simulate these threats, we set up a Wi-Fi AP (monitoring the occurring traffic with \texttt{tcpdump}) and use it as an Internet uplink for off-the-shelf smartphones equipped with SIM cards of commercial operators within our home country.

For invasive traffic-altering attacks, we devised \texttt{iptable} rules that forward the corresponding packets to our MitM (Monster in the Middle) script.
For the traffic rewriting we
reused the Scapy-based implementation of our server-side scanning solution.

\subsection{Outdated Software}
While preparing the exploit chain and testing the setup described above, we identified that Samsung and (some) MediaTek-based devices use strongSwan\footnote{\url{https://github.com/strongswan/strongswan}} as a foundation for their VoWiFi support.

While Samsung uses a recent version of strongSwan (i.e., version 5.9.8 for the Galaxy S24+), 
the \texttt{charon} binary of our MediaTek device (i.e., the Xiaomi Redmi A1) identifies itself to be part of strongSwan 5.1.2, released in March 2014. %

\subsection{Pivoting DH Groups via \texttt{INVALID\_KE}}
\label{sec:pivoting-dh-groups}
\begin{figure}[t]
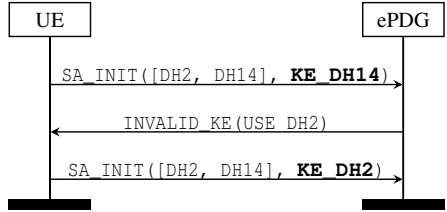

    \centering
    \drawframe{no}
    \setmscvalues{small}
    
    \setlength{\topnamedist}{0cm}
    \setlength{\leftnamedist}{0cm}
    \setlength{\topheaddist}{0cm}
    \setlength{\bottomfootdist}{0cm}
    \setlength{\envinstdist}{0cm}
    \setlength{\levelheight}{0.7cm}
    \setlength{\firstlevelheight}{0.7cm}
    \setlength{\labeldist}{0cm}
    
    \setlength{\instdist}{4cm}
    \scalebox{0.9}{%
    \begin{msc}[msc keyword=]{}
        \declinst{ue}{}{UE}
        \declinst{epdg}{}{ePDG}
    
        \mess{ \texttt{SA\_INIT([DH2, DH14], \textbf{KE\_DH14})} }{ue}{epdg}
        \nextlevel
        \mess{ \texttt{INVALID\_KE(USE DH2)}}{epdg}{ue}
        \nextlevel
        \mess{ \texttt{SA\_INIT([DH2, DH14], \textbf{KE\_DH2})} }{ue}{epdg}
    \end{msc}%
    }%
    \caption{An ePDG server can switch from the initially selected DH group (\DHcmd{14}) to a different group that is offered by the client within the proposal (\DHcmd{02}).}
    \label{fig:ike-invalid-ke}
\end{figure}

Whenever a client connects to an IKEv2 server, it has to communicate its supported SA (Security Association) parameters within the \texttt{SA\_INIT} packet. While it has to decide on a specific key exchange method (i.e., DH group), it can also signal support for other groups within its proposal.
The server can then either accept the proposed key exchange method %
or switch to another offered group by sending an \texttt{INVALID\_KE} (invalid key exchange) message. %
This message can carry the server's proposed method.
The client then retries and sends a fresh \texttt{SA\_INIT} packet with the chosen key exchange, as shown in Figure~\ref{fig:ike-invalid-ke}.
While the proposed SAs within the \texttt{SA\_INIT} packet are normally protected against rewriting attacks by subsequent integrity checks, downgrading by the \texttt{INVALID\_KE} message is possible because the client discards its current state and starts from scratch with the indicated key exchange.

Thus, an active attacker can suppress the first \texttt{SA\_INIT} message and send the client a spoofed \texttt{INVALID\_KE} packet proposing a lower DH group and effectively, downgrading the key length. %
Clients supporting weak DH groups and servers tolerating insecure proposals (without demanding the client to switch to a stronger group if available) facilitate this kind of downgrade attack.

\subsubsection{Results}
Properly implemented clients that propose multiple SAs are prone to this attack. 
For example, since Apple devices only announce a single DH group within their client-side proposal, they are not vulnerable to this kind of downgrade attack.
The same holds true for other scenarios where the client explicitly uses a single DH group (e.g., devices using Samsung's default configuration that is shown in Table~\ref{tab:ike_defaults}).
However, as our client-side analysis showed, most devices are overprovisioned with multiple DH groups, and their settings include deprecated groups. %
For all those devices in our sample, we successfully switched the used key exchange to the weakest offered group using the attack described above -- if they were not already making the weakest selection their default anyway.

In context with our results probing the ePDGs, where we have seen that at least 41\% of the operators tolerate weak client preferences over stronger available DH groups, we can conclude that it is feasible to execute this attack under real-world circumstances.

\subsection{MediaTek Implementation Bug}
\label{sec:mediatekbug}
Besides testing our available UEs for (maliciously played) \textit{protocol-conform} downgrade attacks (as described above), we also tested whether the \texttt{INVALID\_KE} message is properly implemented. %
Specifically, we test how the client will react to the server's (or spoofed) request to switch to a DH group not part of the preloaded configuration.

\begin{figure}[!t]
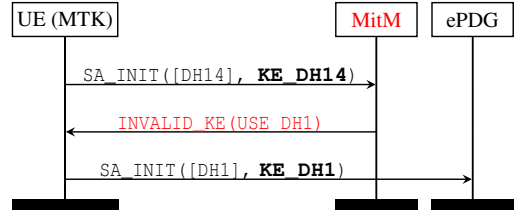

    \centering
    \drawframe{no}
    \setmscvalues{small}
    
    \setlength{\topnamedist}{0cm}
    \setlength{\leftnamedist}{0cm}
    \setlength{\topheaddist}{0cm}
    \setlength{\bottomfootdist}{0cm}
    \setlength{\envinstdist}{0cm}
    \setlength{\levelheight}{0.7cm}
    \setlength{\firstlevelheight}{0.7cm}
    \setlength{\labeldist}{0cm}
    
    \setlength{\instdist}{3.2cm}
    \scalebox{0.9}{%
    \begin{msc}[msc keyword=]{}
        \declinst{ue}{}{UE (MTK)}
        \declinst{mitm}{}{\textcolor{red}{MitM}}
        \setlength{\instdist}{0.2cm}
        \declinst{epdg}{}{ePDG}
    
        \mess{ \texttt{SA\_INIT([DH14], \textbf{KE\_DH14})} }{ue}{mitm}
        \nextlevel
        \mess{ \textcolor{red}{\texttt{INVALID\_KE(USE DH1)}}}{mitm}{ue}
        \nextlevel
        \mess[text width=5cm,align=left]{ \texttt{SA\_INIT([DH1], \textbf{KE\_DH1})} }{ue}{epdg}
    \end{msc}%
    }%
    \caption{Some of MediaTek's Dimensity basebands are vulnerable to severe downgrade attacks, allowing the selection of DH groups that are known to be weak and were never part of the initial \texttt{IKE\_INIT}-proposal nor 3GPP specification.}
    \label{fig:ike-downgrade-mediatek}
\end{figure}

\subsubsection{Results}
While all devices behaved as expected (rejection of the offer), some MediaTek-based devices stood out.

We found that there are at least two different IPsec implementations for VoWiFi support on MediaTek devices.
The first one (presumably older devices with a Helio chipset) uses strongSwan (based on strongSwan-related configurations in the firmware images).
In contrast, newer Dimensity-based devices lack the necessary strongSwan files and thus presumably use a different IPsec/VoWiFi stack.

Our active measurements on multiple models reveal that the latter MediaTek devices are not only \textit{vulnerable} to the attack described above, but also accept \texttt{INVALID\_KE} fixations to any DH group (i.e., also to a group that was not part of the UE's proposal). In practice, an attacker can thus downgrade to the weakest \dh01, as shown in Figure~\ref{fig:ike-downgrade-mediatek}.
According to our threat model and Adrian et al.~\cite{adrian15weakdh}, we consider this breakable by well-funded academic researchers and certainly within reach of resourceful (not necessarily nation-state) actors.
We want to emphasize that the two downgrade attacks described above work on unmodified, unrooted smartphones within commercial networks.

\subsection{Responsible Disclosure}
We disclosed those vulnerabilities to MediaTek and a fix is available (Appendix \ref{sec:responsibledisclosureMediaTek}).

\subsection{Escalating \roundframe{L1} Attacks}%

This Section gives some context, on how \roundframe{L1} attacks can be facilitated to gain control over the full IKE/IPSec/SIP stack. The key observations are:

\paragraph{\roundframe{L1} downgrading drastically eases key recovery} 
Even a downgrade to \dh01 (Section \ref{sec:mediatekbug}) still needs massive computing power to be cracked, but is considered in reach for several years now \cite{adrian15weakdh}. 

\paragraph{Regular Rekeying without Reauthentication}
IKE's keys are regularly regenerated using a DH exchange based on the selected lifetime (see key lifetime analysis in Section \ref{sec:keylifetimes}). 
An attacker can hijack the rekeying as shown in our experiment in Appendix \ref{sec:rekeyingexperiment}. 
However, no authentication is performed on \roundframe{L2} making it roll over into the next session key.

\paragraph{\roundframe{L2} Child SA has no integrity protection} 
It relies on \roundframe{L1} for providing all of the integrity protection.
Since there is no subsequent reauthentication, cracking the outer key exchange is enough to gain stealth rewriting capabilities within the first two layers.

\paragraph{\roundframe{L3} SIP encryption is optional and not enforced}
As shown in Experiment Appendix \ref{sec:SIPencryptionOptional}, SIP encryption is considered optional by most providers.

\subsubsection{Full Attack Outline}
\label{sec:fullL123attackOutline}

\begin{enumerate}%
    \item An active attacker inhibits the first \texttt{INIT\_SA} message from the client to the server and proposes a weak DH group (as described in Section \ref{sec:downgrading}).
    \item From now on, the attacker lets the client and server handshake the weaker DH group, authenticate the connection (via EAP-AKA), and create a session key for \roundframe{L2}.
    \item The attacker can now race to crack the outer key exchange and thus gain rewriting capabilities on \roundframe{L1} -- before the key lifetime expires and a rekeying is triggered. Note: the attacker does not have \roundframe{L2}'s session key yet.
    \item If or when the time comes\footnote{The standard allows both peers to trigger a rekeying prematurely, but we have not tested that.} (Section \ref{sec:keylifetimes}) for a rekeying of \roundframe{L1}, the attacker can handshake both sides independently and inject themselves in between.
    \item Similarly, when \roundframe{L2} is rekeyed, the attacker can handshake both sides independently and inject themselves in between without needing authenticating.
    Note: The attacker now has also control over \roundframe{L2}.
    \item As encryption of the SIP connection \roundframe{L3} is optional (Appendix \ref{sec:SIPencryptionOptional}), an attacker also likely gains control over the client' authenticated IMS session. 
\end{enumerate}

\section{Related Work}
\label{sec:relatedwork}

\paragraph{Encryption in Cellular Networks}
Cryptographic problems have plagued cellular networks from the start. More recently,  Yomna et al.~\cite{yomna2023cellular}
presented Android's approach to combat so-called \textit{null ciphers}.
Null ciphers are mock ciphers that can be inserted into the encryption stack in case no actual encryption is desired. 
Cholesta et al.~\cite{chlosta2019lte} put European networks to the test - many of them still accepted null ciphers on the radio layer.
Tsay and Mj{\o}lsnes~\cite{Tsay2012Vuln} found impersonation vulnerabilities in the Authentication and Key Agreement Protocols (AKA) in UMTS and LTE.
Rupprecht et al.~\cite{Rupprecht2018SecurityResearch} categorized past cellular network vulnerabilities to identify classes of errors and how to combat them.

\paragraph{Diffie-Hellman Groups and IKE}

In \textit{Imperfect Forward Secrecy}, Adrian et al.~\cite{adrian15weakdh} show all the little ways in which DH implementations fail in practice. 
Bhargavan et al. \cite{bhargavan2016downgrade} try to answer the question of how to support reconfigurability while at the same time guaranteeing the preferred mode is negotiated.
Felsch et al. \cite{felsch2018dangers} reports on Bleichenbacher attacks on IKEv1 and IKEv2.

\paragraph{Evaluating Real-World Security Configurations}
Hue et al.\cite{Hue2011Enterprise} evaluated both client- and server-side WPA2-enterprise configurations for education institutes (e.g., eduroam), uncovering deprecated settings and suspected private key sharing across different institutes.
Valenta et al. \cite{Valenta2018CurveSwap, valenta2018search} performed Internet-wide scans for TLS, SSH, and IPsec, surveying their elliptic curve usage and improper curve validation.
Heninger et al.~\cite{Heninger2012Mining} analyzed the occurrence of weak (factorable) keys in the wild.

\paragraph{Evaluation of VPN Servers}
Maghsoudlou et al. \cite{Maghsoudlou2023Characterizing} executed Internet-wide scans to discover and fingerprint VPNs, finding over 7 million IPsec servers.
Kahn et al. \cite{Khan2018Empirical} and Ramesh et al. \cite{rameshvpnalyzer} made large-scale measurements in the commercial VPN ecosystem exposing leaked user traffic.
Wu et al. \cite{Wu2023Back} investigated academic VPNs, which have become an integral part of the home office life.

\paragraph{Roaming Experiments and Large-Scale Cellular Measurements}
Sahin and Francillon~\cite{Sahin2016Over} observed hijacked and thus monetized voice calls being redirected to over-the-top (OTT) services (e.g., WhatsApp, Viber).
Gegenhuber et al.~\cite{Gegenhuber2023MobileAtlas, Gegenhuber2022Zero} introduced a measurement platform enabling scalable cellular measurements by tunneling the communication between SIM card and modem over the Internet.
Besides measurements on the radio layer, Gegenhuber et al.~\cite{gegenhuber2024geoblocking, Gegenhuber2024Never} also evaluated global \mbox{VoWiFi} deployments, exposing geoblocking practices at \mbox{VoWiFi} by simulating clients from different countries.\

\paragraph{SIP in VoLTE and RCS}
Tu et al.~\cite{tu2016new} uncovered spoofing and injection vulnerabilities at VoLTE's SIP layer.
Similarly, Yang et al.~\cite{yang2024uncovering} exposed weaknesses in real-world RCS deployments.
In 2023, the Google Project Zero team discovered four severe Exynos vulnerabilities, including a remote code execution on the most recent Pixel and many Samsung baseband processors by injecting malicious SIP messages~\cite{Google2023Exynos} into the VoLTE/VoWiFi traffic.%

\section{Discussion}
\label{sec:discussion}

This paper set out to cartograph the state of \vowifi on the UE and MNO side. Little did we know what awaited us. 
The ecosystem is haunted by multiple structural and standardization problems: 
\begin{enumerate}%
    \item[a)] 
    an inadequate and slow process of provisioning provider settings to the UEs, with too many middlemen, %

    \item[b)] structural disincentives for phasing out %
    deprecated cryptography and a na\"ive standardization approach,

    \item[c)] optional encryption in certain parts of the ecosystem and the prevalence of the dumb client paradigm,

    \item[d)] critical bugs on the UE and MNO side.
\end{enumerate}

\subsection{Provisioning and Configuration}
The missing VoLTE/VoWiFi autoconfiguration feature inspired handset manufacturers to find (non-interoperable) ways to preload known settings and curate their own databases \roundframe{RQ1}. It is a painful, tedious task for the operators and the handset manufacturers alike, with multiple middlemen that do not inspire quick, painless updates to new settings, disincentivizing updates. 

This is represented in the very inconsistent settings among the different vendors (Section \ref{sec:staticUEconfig}) and the large adoption of deprecated DH groups in provider-specific settings (Section \ref{sec:activemnoscanning}) as well as default settings, as seen in Figures \ref{fig:dhgroups} and \ref{fig:ike_deprecated}. 

Ultimately, MNOs should have the power to make configuration changes, including removing deprecated cryptographic algorithms without impairing service, if they wish to.

\subsection{Structural Hurdles of Deprecation}
\label{sec:discussionDeprecation}
The MNOs have little to no incentive to phase out older insecure key exchange methods \RQ2. %
On the one hand, (anticipated) compatibility issues with legacy devices and the slow update process might stoke sentiments against changes. %
In our sample, only 7\% of operators ditched all the insecure DH groups below 2048 bits.

On the other hand, 3GPP/ETSI lacks a defined depreciation path. Just removing it from the standard does not actually remove the method from the world nor the affected devices. %

In standards, there is no room to be stingy on the number of key bits. If anything, it is the place to be bold and visionary. 
If shorter lengths are required at the start, a stringent phase-out plan/process should be defined with it. The development of computing power turned out to be somewhat predictable, %
and the same can be expected for the deprecation of key lengths.

\subsection{Optionality and Strict Configurations}
The \textit{dumb client} paradigm, often found in large infrastructure, envisions the majority of decisions to be made by the network and not the client. 

\textit{Using SIP encryption? If the network does not mandate it, the client will definitely not object.} 

However, the VoWiFi ecosystem, which is built upon many Internet technologies, has the infrastructure and protocolary means for clients to request better settings at their discretion. UE chipset and operating system manufacturers should take this chance.

Furthermore, the data suggest the importance of those preloaded configurations might also be simply overestimated - as seemingly conflicting carrier configurations from different handset vendors still work, and 42\% of the operators request an upgrade of the key exchange method if a common higher group is available. 3GPP, the operators, and handset/baseband manufacturers should trust more in autoconfiguration measures (or enforce their own minimum standards), even at the expense of slightly longer connection times.

\subsection{Downgrades, Bugs, and Vulnerabilities}

Attacks (Section \ref{sec:fullL123attackOutline}) against \dh01 still require heavy lifting for cracking the key exchange within the key lifetime, but it is assumed to be within reach for resourceful attackers. This is not for everyone, and nation-state actors would, therefore, likely choose a legal approach for domestic key seizure. 

However, downgrading to a weaker DH group alone should already be considered a serious vulnerability. Otherwise, selecting different key lengths would be pointless.

Recovered (downgrade attack) and leaked static keys do not always have to be used to attack higher layers up to the SIP/IMS connection (snooping conversations or spoofing commands). \roundframe{L1} decryption alone can be used as a type of IMSI Catcher\cite{Dabrowski2014IMSICatcherCatcher} on VoWiFi by sniffing EAP-AKA identifiers. 

\subsubsection{IPsec Rekeying Problems}
We see a number of downgrade attacks against the IKE phase~1 manifesting in the 3GPP VoWiFi ecosystem \RQ{3}. An attacker should not be able to force an exchange method and bit length upon the parties. 

Attacks on \roundframe{L1} and the rekeying system gain impact because they inherit a previous EAP-AKA authentication on \roundframe{L2}. And since \roundframe{L3} SIP Encryption is optional in many cases (Experiment see Appendix \ref{sec:SIPencryptionOptional}), this gives an attacker control over all three signal and user data panes. 

\subsubsection{Accepting Undocumented and Unoffered Algorithms or Key Lengths}
The MediaTek vulnerability of accepting unproposed and non-complaint key exchange methods is an \textit{insecure implementation} by over-fulfilling the specification, as described by Rupprecht et al.~\cite{Rupprecht2018SecurityResearch}. 
In this type of implementation error, the attack surface is unnecessarily enlarged (the client accepts a larger input language than required or even advertised) by including extra functionality outside of the specification.
The weak key exchange method was likely inherited from a general-purpose IKE implementation or library. The developer removed it from the advertised methods but never checked the received selection. Ideally, unsupported methods should also be removed from the code.

\subsubsection{Not-so-private Private Keys}

As the manufacturer was identified as the source of the global static set of ten round-robin keys, they can not be considered \textit{private} for a number of reasons:
\begin{itemize}%
    \item[a)] All of the affected operators are in possession of the same keys and can decode each other's traffic.
    \item[b)] The manufacturer (and any demo or test customer) are also in possession of those keys.
    \item[c)] Security or private actors could seize the opportunity to get those keys from an institution under their jurisdiction or other control. 
    \item[d)] Used telco equipment finds its way to second-hand hardware marketplaces \cite{Schmitt2015Attack} and might leak those keys into the public.
\end{itemize}

\section{Evasive Recommendations}
\label{sec:recommendations}

\subsection{Default to Strongest DH Group}
Operator configurations that are preloaded to clients are updated only irregularly and thus often outdated.
In practice, supported DH groups within those configs are often comparatively weaker than on the server side.
To counter this, clients should treat the preloaded options as a lower bound, and always signal (and prefer) stronger DH groups in their proposal.
In the worst case, this adds another roundtrip where the server indicates that it does not support that mode. To save that on subsequent connections, server capabilities could be temporarily cached. %

\paragraph{Failure Mode} A rejected VoWiFi handshake on security grounds does usually not lead to loss of service for the customer, as the phone falls back to cellular service.

\subsection{Not-so-private Private Keys}
Leakage or re-usage of private keys can happen for a number of reasons - but from the perspective of a phone that most of the time connects to a single operator's \epdg, only the intra-operator reusage is detectable, not the inter-operator reusage. 

\paragraph{UE-local Freshness Tests}
In lieu of a cryptographically ensured freshness, the client can detect key intra-operator re-usage with a history mechanism. However, based on the observation time frame and the network volatility, this history might grow large. 
A constant size and complexity key history could employ a temporal ring of Bloom Filters \cite{dabrowski2016mobile}.

\paragraph{Distributed methods}
Inter-operator key re-usage detection would require cooperation between a (vast) number of phones either with a common infrastructure (e.g. like DNS blocklists) or in peer-to-peer mode.

\paragraph{Failure Mode}
A security-rejected handshake is tolerable, as the user would not experience a loss of service due to fallback to cellular service. 

\subsection{Fallback to an Unannounced Mode}%

\textit{Do not roll your own crypto!} is valuable advise. 
However, if a standard library is used, unsupported methods and ciphers should be removed not only from the negotiations but also from the code base. 
A missing test coverage for a particular piece of code could either hint at a missing test or a removable over-implementation.

\subsection{Defined Upgrade Path in Standardization}
As discussed in Section \ref{sec:discussionDeprecation}, standards of cryptographic applications should define an upgrade timeline for minimum supported security features, such as key length.

\section{Conclusion}
\label{sec:conclusion}

The VoWifi ecosystem relies on IKE and IPsec to set up secure tunnels into the operator's EPC. 
However, multiple factors lead to a delayed adoption of up-to-date key exchange mechanisms. Deprecated DH groups (by 2015 standard) and other dated cryptographic primitives are the norm on the client and the operator sides in 2024 -- and computing power only got cheaper in that time frame. 

Furthermore, we encountered client implementation issues with a major %
smartphone SoC vendor, facilitating downgrade attacks to weak, non-compliant key exchange methods.

The biggest surprise was the operator side, as at least 13 operators serving 140 million customers apparently used the same global set of static private keys.
In both cases, we helped to remove those vulnerabilities through responsible disclosure programs and tracked their progress.

\section*{Acknowledgments}

This paper has been in part funded by the \textit{UniVie Doctoral School Computer Science} and \textit{Christian Doppler Laboratory for Security and Quality Improvement in the Production System Lifecycle} with financial support by the Austrian Federal Ministry of Labor and Economy, the National Foundation for Research, Technology and Development and the \textit{Christian Doppler Research Association}. %
Further support by the \textit{Usable Security Group} led by Katharina Krombholz at the\textit{ CISPA Helmholtz Center for Information Security}.
Further funding originates from the \textit{NGI0 Entrust Fund}, established by \textit{NLnet} with financial support from the European Commission's Next Generation Internet, %
under grant agreement No 101069594 and Project FO999887504 \textit{DynAISEC} funded by the
Program \textit{ICT of the Future}. %
The competence center \textit{SBA Research} (SBA-K1) is funded within the framework of \textit{COMET – Competence Centers for Excellent Technologies} by BMVIT, BMDW, and the federal state of Vienna, managed by the FFG.

\bibliographystyle{plain}
\let\oldbibliography\thebibliography
\renewcommand{\thebibliography}[1]{%
  \oldbibliography{#1}\small%
  \setlength{\itemsep}{0pt}%
  \setlength{\parskip}{2pt plus 3pt}
}
\bibliography{base, rfc, subscriber-sources}
\normalsize
\appendix

\section*{Appendix}
\label{sec:appendix}

\section{Experiment: SIP Encryption Optionality}
\label{sec:SIPencryptionOptional}
Not all operators enforce encryption and integrity on \roundframe{L3} in practice, which leaves room for even more severe attacks \roundframe{G1-3}.
This is partly visible in our static configuration file analysis but also in-vivo verifiable:

For example, when we removed client-side encryption and authentication preferences on our testing device (using the stock MTK EngineerMode app), we were still able to connect to the home operator successfully.

In such cases, an attacker that cracked the outer IKEv2 key exchange and is thus able to take over the first two layers can subsequently also hijack the third layer after the SIP authenticated between UE and P-CSCF is finished, effectively dominating all three communication layers and reaching all available goals~\roundframe{G1-3}.

\begin{table}[!t]%
    \centering%
    \caption{Deprecated IKEv2 SA proposal parameters \cite{etsi-ts-133.210,rfc8247,rfc9395}}%
    \label{tab:deprecated}%
    \resizebox{0.8\linewidth}{!}{%
    \begin{tabular}{l l l }
    \toprule
        Category & ID & Name \\
    \midrule  
    Encryption Algorithms & 1 & DES IV64 \\
       & 2 &  DES  \\
       & 4 &  	RC5  \\
       & 5 &  IDEA \\
       & 6 &  CAST \\
       & 7 & BLOWFISH \\
       & 8 & 3IDEA \\
       & 9 & DES IV32 \\ \hline
    Pseudo-Random-Functions & 1 & HMAC MD5 \\      
      & 3 & HMAC Tiger \\ \hline
    Integrity Algorithms &  1 & HMAC MD5\_96 \\ 
       &   3 & DES MAC \\
       &   4 & KPDK MD5 \\
       &   6 & HMAC MD5\_128 \\
       &   7 & HMAC SHA1\_160 \\ \hline
    Key Exchange Methods & 1 & 768 Bit MODP \\
        & 2  & 1024 Bit MODP \\
        & 5 & 1536 Bit MODP \\
        & 22 & 1024 Bit MODP 160 Prime \\
    \bottomrule
    \end{tabular}
    }%
\end{table}

\section{Experiment: No Integrity Protection in Regular Rekeying}
\label{sec:rekeyingexperiment}

As described in Section \ref{sec:pivoting-dh-groups}, downgrade attacks via the \texttt{INVALID\_KE} message are not integrity protected and work as a stepping stone to taking over the full \roundframe{L1-L3} stack. 

In this experiment, we verify that the same is true for the regular rekeying of \roundframe{L1}. 

In this instance only, to simulate the capability of breaking the key exchange, we used a rooted phone.
We inject Frida\footnote{\url{https://github.com/frida/frida/}} into the process responsible for the IKEv2-related communication and extract the used encryption and authentication keys by intercepting the corresponding library functions.

Thus, we were able to manipulate the rekeying interval and observe it in vivo. 

The results confirm that regular rekeying lacks integrity protection similar to \texttt{INVALID\_KE}-triggered rekeying.

\section{Static Key Re-usage: Mapping MCC-MNC to Operators}
\label{sec:mnocounting}
As mentioned in Section \ref{sec:MethodologyLimitationsMapping}, an MCC-MNC tuple does not necessarily constitute an operator. Old MCC-MNCs are often kept alive for historical reasons. 

In our set of 15 ePDGs using static keys (Table \ref{tab:remediation}), three pairs had identical IP addresses. \textit{Hutchison Drei} actively maintains 232-05 and 232-10 after merging \textit{Orange} and \textit{One}. The case is very similar for \textit{Smartfren} and their 510-09 and 510-28 designations. In contrast, Malaysia's  \textit{U Mobile} and \textit{DiGi} cooperate by maintaining a common 5G infrastructure but are otherwise (mostly) independent operators. Thus, the latter ones are counted as two operators.

Pakistan's Telenor newly showed up in our scans on April 2nd, 2024, over a month after we started the responsible disclosure, and several operators had already rolled out the patch.

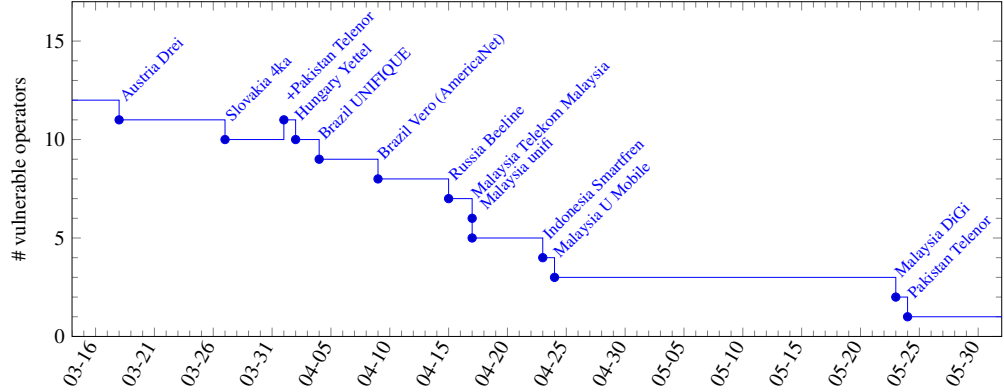
\begin{figure*}[!t]%
\centering%
    \resizebox{0.75\linewidth}{!}{%
    \begin{tikzpicture}%

  \begin{axis}[date coordinates in=x,date ZERO=2024-03-01,
               xticklabel=\month-\day,
               xticklabel style={rotate=60,anchor=east},
               minor x tick num=4,
               minor y tick num=4,
               ymin=0,ymax=17,
       xmin=2024-03-14,
       xmax=2024-06-01, 
       ylabel={\# vulnerable operators},
       x=2.0mm]

\addplot+[const plot mark left,
    every node near coord/.append style={xshift=-1pt,yshift=8pt,anchor=west,font=\footnotesize,rotate=45},
    nodes near coords={\labelz}, 
    visualization depends on={value \thisrowno{2}\as\labelz}] 
table[col sep=comma,x=date,y=num]
{
date,    num, name,    mccmnc
2024-01-01,12,,
2024-03-18,11,	Austria Drei,	232-05 232-10
2024-03-27,10,	Slovakia 4ka,	231-03
2024-04-01,11,	+Pakistan Telenor,	410-06
2024-04-02,10,	Hungary Yettel,	216-01
2024-04-04,9,	Brazil UNIFIQUE,	724-29
2024-04-09,8,	Brazil Vero (AmericaNet),	724-26
2024-04-15,7,	Russia Beeline,	250-99
2024-04-17,6,	Malaysia Telekom Malaysia,	502-11
2024-04-17,5,	~~~Malaysia unifi,	502-153
2024-04-23,4,	Indonesia Smartfren	,510-09 510-28
2024-04-24,3,	Malaysia U Mobile,	502-18 
2024-05-23,2,	Malaysia DiGi,	502-16
2024-05-24,1,	Pakistan Telenor,	410-06
2024-12-31,1,,
};
 
\end{axis}
\end{tikzpicture}%
    }%
    \vspace{-1ex}\caption{Globally Static Set of DH Keys: Remediation over Time}%
    \label{fig:remediation}%
    \vspace{-1ex}%
\end{figure*}

\section{Responsible Disclosure and Remediation}
\label{sec:responsibledisclosure}
\subsection{MediaTek Unannounced DH Group and Downgrade}
\label{sec:responsibledisclosureMediaTek}
MediaTek confirmed our findings and issued CVE-2024-\censor{20069}\footnote{\url{https://corp.mediatek.com/product-security-bulletin/June-2024\#CVE_2024_20069}} (severity: high) for the described downgrade attack. 
The affected basebands\footnote{MT6833, MT6853, MT6855, MT6873, MT6875, MT6875T, MT6877, MT6883, MT6885, MT6889, MT6891, MT6893, MT8675, MT8771, MT8791T, MT8797} with the NR15 modem are from the Dimensity product line. 

They released patches to all affected customers.
All Android devices with a Security Patch Level (SPL) of 2024-06-05\footnote{\url{https://source.android.com/docs/security/bulletin/2024-06-01}} or later are protected from the downgrade attack.

\subsection{Globally Static Set of DH Exchange Keys}
\label{sec:responsibledisclosureGSMA}
After the experience with the few and slow responses from the operators themselves~\censor{\cite{Gegenhuber2023MobileAtlas}}, this time we reached out to the GSMA's Coordinated Vulnerability Disclosure (CVD) program to timely contact the affected operators and the manufacturer on 2024-02-13. We further reached out to Apple and Google to consider countermeasures for their mobile operating systems.

The GSMA has issued CVD-20\censor{24-0089} to track our findings and further helped to communicate them with the affected manufacturers and operators.

ZTE confirmed our findings and issued CVE-2024-\censor{22064}\footnote{\censor{\url{https://support.zte.com.cn/support/news/LoopholeInfoDetail.aspx?newsId=1035524}}} (severity: high). %
The software component responsible is ZXUN-ePDG from their CCN (Computing and Core Network) product line.
According to ZTE, the bug has been present in all versions before V5.20.20.
The issue was caused by incorrectly shipping integration test keys in the production release, they explained.
Besides a fixed version, ZTE also offers a volatile runtime-only fix for MNOs that can not currently be updated; it must be reapplied after each restart.

\begin{table}[t]%
\caption{Static IPSec keys: Vulnerable Operators.}%
\label{tab:remediation}%
\newcommand\tFa{$^{\mathrm{a}}$}%
\newcommand\tFb{$^{\mathrm{b}}$}%
\newcommand\tFc{$^{\mathrm{c}}$}%
\resizebox{\linewidth}{!}{%
\begin{tabular}{@{}lll@{}c@{}r@{}r@{}}
\toprule
MCC-MNC        & Country   & Operator           & \clap{Subscribers(M)} &        & Remediation\tFb \\ \midrule
232-05, 232-10 & Austria   & Drei              & 4.1&\cite{Sub23205}    & 2024-03-18  \\
231-03         & Slovakia  & 4ka               & 0.6&\cite{Sub23103}    & 2024-03-27  \\
216-01         & Hungary   & Yettel            & 3.7&\cite{Sub21601}    & 2024-04-02  \\
724-29         & Brazil    & UNIFIQUE          & < 0.5&\cite{Sub724xx}  & 2024-04-04  \\
724-26         & Brazil    & Vero (AmericaNet) & < 0.5&\cite{Sub724xx}  & 2024-04-09  \\
250-99         & Russia    & Beeline           & 44&\cite{Sub25099}     & 2024-04-15  \\
502-11         & Malaysia  & Telekom Malaysia  & 2&\cite{Sub50211}      & 2024-04-17  \\
502-153        & Malaysia  & unifi             & 0.8&\cite{Sub502153}   & 2024-04-17  \\
510-09, 510-28 & Indonesia & Smartfren         & 36&\cite{Sub51009}     & 2024-04-23  \\
502-18         & Malaysia  & U Mobile          & 8.5&\cite{Sub50218}    & 2024-04-24  \\ %
502-16         & Malaysia  & DiGi              & 20.6&\cite{Sub50216}   & 2024-05-23  \\
410-06\tFa    & Pakistan  & Telenor           & (44)&\cite{Sub41006}      & 2024-05-24  \\ 
429-01         & Nepal     & Nepal Telecom     & 20&\cite{Sub42901}     & - \\
\midrule
Total          &           &                   & > 140.3                &  Mio     &      \\ \bottomrule
\end{tabular}%
}%
\newline\footnotesize
\tFa~Vulnerability introduced April 2nd 2024.\hspace{1ex}
\tFb~Cut-off date: May 31th 2024
\end{table}

\subsubsection{Remediation Timeline} 
To track the progress of the update, we ran scans hourly. 
In mid-March 2023, one operator confirmed to us that they received a patch and were testing it. 
Shortly thereafter, most operators started rolling out the patch into production. 
Figure \ref{fig:remediation} and Table \ref{tab:remediation} note the last time for each operator, we recorded an ePDG server using one of those static keys.

Interestingly, Pakistan's Telenor first started showing up as vulnerable on April 2nd in the scans well after a fix was available. We were later informed, that this is a test site not ready for commercial use and will be upgraded before it is open to customers.

\vspace{3ex}

\subsubsection{Key Hashes}
We include the SHA256 hashes of the found globally-used non-random public keys (bytes in network order) to facilitate blacklisting of those keys. Because of space constraints, we only include \dh01 through \dh15. 

\vfill\eject

\scriptsize\DHcmd{01}:\newline
\tt\scriptsize
c91fbb17c38e95c3590c54838bab62808df808cce198c3ba24e830c8f3cc2fc7  \newline
564a3bad4f7504c4c1c515b8cad7687cfdf19af0cce3b527cfc56093fab266c8  \newline
7050ed52e9222665ad11f20cc51218c253a1a54695ab246bf0e408d9ac041176  \newline
498f3bbfa8f8f2b5b1b3ec7cd6790d960f0c760ecbaf5eca5d31752aa2fcc1fd  \newline
0caf2731a9dfd392821134adff2f0ffa8097e220dcd0955c3370571da0ee9586  \newline
2ade9b103ce1cb75a7894905b55c51ec5cf6bc78513cdd9373c3266f0d1c2ed2  \newline
4612bc13632fb814fac5ce1b24aba1a79ef8284ea737b241e5311423c0510782  \newline
43090ffd7d3285678d3b1d98e4bfbe5775911a5595258639287a641b48eb32a3  \newline
b2b8e7eb53256495519209eebd98a2f9d7974241c848c7ad37ec586676ae4116  \newline
a005b8aac47ce34c6293f1f37da868107f5e05db5aebb4618d899a94927af47e  
\rm

\DHcmd{02}:\newline
\tt\scriptsize
e99889815a602ab1863e4d900650233bfc00aa64ce29fde18c36219f6aa361e7  \newline
a9f58f41c2e3b1d36f74d14e0a60e7e833e1faf438b71e42a0ee76fd208420d3  \newline
d24c415adb25bb2a8adbde6ee9da4f5c65c39b746a87e9ac71b613b664720c41  \newline
9c3fee28e4a984db043924ccd42a8121bf1fd696428a82fac624197df10a4d35  \newline
3289fe1551fc0fbf372f293ffa9867c52e30d2357d3ab2ae5d4b8c896f4b57b5  \newline
9dd792e808954c00fc1565a447324788a34913a7df977a836bd523edca1a89d5  \newline
e8bc246f549cab69d4e6789cfa611d24828d532f839597c02d7b193ae0dc7cab  \newline
423516e9e2e67f0018abf66e20f2ded682da51d12752fe010103698782575f6f  \newline
936aeb8c8d413d03682a0ab68ed7ae0f98c0be630575704eb3f395d70bea83ab  \newline
01870a3a8e23257f81ea50e84a7cac4d7be949c18681725576c66e32b811c6b9  
\rm

\DHcmd{05}:\newline
\tt\scriptsize
b22cdb284a4db637fe76f1f7ab8a1f8c2530976805822686b008176e60e5cf29  \newline
398c99f3e84bef849fc62a6216c5b66f95445a92ce6b8d49b56f8c73c994e774  \newline
578a16523d85ac4ab566b7cefa2545a64a4a7175b4432eab9a2a5e4200a3e391  \newline
74e0aa079e655f8caeefa0dc35c8c81d5c81e4bc4b5d3e4f975a1f1ed198f68f  \newline
d083fb11ed987a56f28e5887980ead2a173396f0676e5b322aa58e81dd85d4c5  \newline
2ba8036874c1cc870d280d1a388a2746ac6ed62f26e427362c8aadc7e2ea13bd  \newline
337095ee22be46044dd0e4626b0b0a19728273001bda8fbf7e77afc514fc8e5f  \newline
4a1d84aab699e209fd96dd611f3c25128c314d37b43fa4d325057b7a1943f09a  \newline
9bd20704d7ab4c879514f2d69f2cbcd8787cbed3c44db0843fe6c540de143799  \newline
662cf2600a8432a09a96f7e2ca480df04712c5ef58d51c95c4341bb085a80491  
\rm

\dh14:\newline
\tt\scriptsize
9247ccc73f6fcc832fd43458f96c8517d45df5548d9841650aec23dae765d918  \newline
c75706b089bfc671a7678661786f6aaad53afa5f57d415d305e0d7e4ee996c0b  \newline
8f9dd2cd05e0b1884e9dde3dcc74b01d782014df68b86bc074782a7a2d78b467  \newline
5aaa0f715a5affea5b40e1e8dad902d8f90adb64b02a99f36b01340878f39a89  \newline
55895ae426a8487ccde1a38cc6631e5728a5eb8605269d4c907f2a93c09e1e03  \newline
e0b4747bff0898e52ef435e930842dea8f3a48ca7d0e3e37b32a0e390a81adaf  \newline
b156c9089f74e32a2e88e110b24010aa1824c5ec625d591081f8d46c49f4b7d2  \newline
08e81e262766a1baadde5ad543a83477be7153394d7dba6d61682eb3f8e24284  \newline
58bfb525033d093dfcce2c483c25495dad2591965253cfbddfc98579c6e8bdf4  \newline
59b6b9094b1b6dfd637d80cc59c8a44160e053c0bb1c66e1f58e1a871258a53a  
\rm

\dh15:\newline
\tt\scriptsize
13ca255b94a7284399177e828f1f39c4a66d618cd735455e5391b4445c603c8d  \newline
5a4f387f10bee59a6209244e43d0eaa67c1e6255c5b2376ffce51f7448d72870  \newline
9ce716182a2790cebe900630bdef64def59ed90e45e7d87029b60d145c20a22b  \newline
1f57f25e95eaeba86e5004b03058433378367e5db9126483b10b9a9262cd25b1  \newline
855aa6a8bb2b2327f52ed5d791f7ef211cc3177e50fa47c907f9a9004e91d002  \newline
b0b84567c90008babed048914325b15ff016d72b5aa46283447a6ea0b16a8fe5  \newline
8fdc2945a14dd9e7f6646107b9d324ab16a5d378e138c282c679f1de343fe447  \newline
516fb8ca4462ff067cd0611f391a289d1aaae6e73b2d96a5d7ad8444ce714a6f  \newline
a9fc5fabd3b4d94c2d11eb4ece812548a93ecb87a4ce82f883b55e6f683a5bc8  
\rm
\newline

\normalsize

\end{document}